\documentclass[aps,prb,twocolumn,showpacs]{revtex4}
\usepackage{graphicx}
\usepackage{amsmath}
\usepackage{amssymb}
\usepackage{fbpack}
\usepackage{color}
\usepackage[normalem]{ulem} 
\usepackage{multirow}
\usepackage{array}
\usepackage{upgreek}

\newcommand{\hz}{\hat h^z}

\newcommand{\vecI}{\hat {\mvec I}}
\newcommand{\Ikx}{\hat I_k^x}
\newcommand{\Iky}{\hat I_k^y}
\newcommand{\Ikz}{\hat I_k^z}

\newcommand{\Sz}{\hat S^z}
\newcommand{\Spl}{\hat S^+}

\newcommand{\sx}{\hat \sigma^x}
\newcommand{\sy}{\hat \sigma^y}
\newcommand{\sz}{\hat \sigma^z}

\newcommand{\ks}{{k_s}}

\begin{document}


\title{Enhanced hyperfine-induced spin dephasing in a magnetic-field gradient}


\author{F\'elix Beaudoin}
\affiliation{Department of Physics, McGill University, Montr\'eal, Qu\'ebec, Canada H3A 2T8}
\author{W.~A.~Coish}
\affiliation{Department of Physics, McGill University, Montr\'eal, Qu\'ebec, Canada H3A 2T8}

\date{\today}

\begin{abstract}
	Magnetic-field gradients are important for single-site addressability and electric-dipole spin resonance of spin qubits in semiconductor devices. We show that these advantages are offset by a potential reduction in coherence time due to the non-uniformity of the magnetic field experienced by a nuclear-spin bath interacting with the spin qubit. We theoretically study spins confined to quantum dots or at single donor impurities, considering both free-induction and spin-echo decay. For quantum dots in GaAs, we find that, in a realistic setting, a magnetic-field gradient can reduce the Hahn-echo coherence time by almost an order of magnitude. This problem can, however, be resolved by applying a moderate external magnetic field to enter a motional averaging regime. For quantum dots in silicon, we predict a cross-over from non-Markovian to Markovian behavior that is unique to these devices. Finally, for very small systems such as single phosphorus donors in silicon, we predict a breakdown of the common Gaussian approximation due to finite-size effects.
\end{abstract}


\maketitle

\section{Introduction}

One of the most intriguing features of spins in semiconductor nanostructures is our ability to systematically determine decoherence mechanisms for these systems from microscopic models. Although complex spin dynamics in the presence of the relevant interactions can be difficult to evaluate, these interactions can be known to a high degree of precision and irrelevant terms systematically dropped from the Hamiltonian.  Indeed, for electron spin qubits confined to quantum dots in GaAs or natural Si, the dominant contribution to dephasing is known to be the Fermi contact hyperfine interaction between the electron spin and nuclear spins in the surrounding host material.\cite{PhysRevLett.100.236802,bluhm2010dephasing,PhysRevLett.88.186802,maune2012coherent}

The contact hyperfine interaction divides naturally into two contributions: a secular (diagonal) part that commutes with the electron Zeeman term, for which dynamics are reversible by Hahn echo,\cite{bluhm2010dephasing} and a non-secular (off-diagonal) part, the so-called flip-flop interactions, which may lead to a more complex dynamics.  In the presence of a large electron Zeeman splitting compared to the total hyperfine interaction strength, perturbative expansions can be used to calculate the contribution of flip-flop interactions to decoherence.\cite{PhysRevB.77.125329,PhysRevB.81.165315} Alternative techniques such as cluster expansions or high-order partial resummations have also been used to estimate dynamics due to flip-flops in the low-field limit.\cite{PhysRevB.74.035322,PhysRevLett.102.057601,PhysRevB.79.245314,PhysRevB.82.035315,Barnes2012} 

In addition to hyperfine interactions, dipole-dipole coupling between nuclear spins can induce internal dynamics in the nuclear-spin bath, which can act back on the electron spin through hyperfine interactions, resulting in decoherence.  This process of electron-spin decoherence due to fluctuations in the nuclear-spin bath (spectral diffusion \cite{Klauder1962}) is independent of the applied magnetic field for moderately large fields and is not fully reversible by spin-echo,\cite{Witzel2005,yao2006theory} thereby typically limiting achievable coherence times.

Electron spins confined in silicon using either quantum dots\cite{maune2012coherent} or at single phosphorus donors\cite{morello2010single,pla2012single} may have significant advantages over spins in GaAs quantum dots, since natural silicon contains only $4.7\%$ nuclear-spin-carrying $^{29}$Si, a figure that can be further reduced by isotopic purification.

Although the long coherence times achievable for semiconductor spin qubits are certainly an advantage, control and scalability are also required for potential applications in quantum information processing. A useful tool in this respect is a magnetic-field gradient. Indeed, electron-spin resonance of single electron spins in lateral dots in GaAs has been demonstrated using inhomogeneous magnetic fields generated by cobalt micromagnets.\cite{pioro2008electrically,PhysRevB.81.085317} Recent proposals suggest to use these gradients to achieve strong coupling of single electron spins to a coplanar-waveguide resonator,\cite{PhysRevB.86.035314} in particular in Si where coherence times are longer. Interactions with a resonator can be used for a qubit readout\cite{PhysRevLett.108.046807,petersson2012circuit} or to mediate interactions between distant spin qubits,\cite{PhysRevB.77.045434} an important step toward scalability. 

\begin{figure}
	\includegraphics[scale=1]{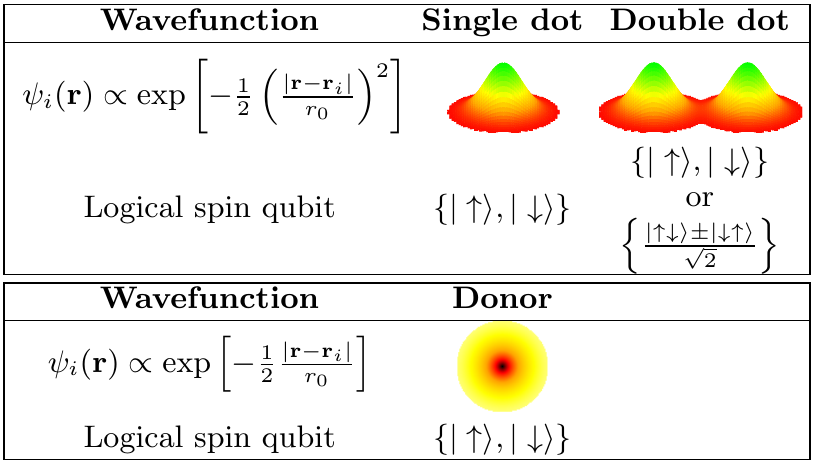}
	\caption{(Color online) Various physical systems considered in this paper. We study spin dynamics for both single- and double-dot geometries. In the double-dot case, we consider either (i) a single electron shared by the two dots or (ii) an electron in each dot, with spin states expressed in the singlet-triplet basis, $\{\ket{S}=(\ket{\upa\dwna}-\ket{\dwna\upa})/\sqrt2$, $\ket{T_0}=({\ket{\upa\dwna}}+{\ket{\dwna\upa}})/\sqrt2\}$. The associated electron envelope wavefunctions can be either hydrogenic (relevant for a donor impurity) or Gaussian (relevant for a quantum dot) in dimension $d=1, 2,$ or $3$. \label{fig:0}}
\end{figure}

In this article, we calculate the effect of such an inhomogeneous magnetic field on the coherence of (electron or heavy-hole) spin qubits.  These results can be applied to a number of experimentally relevant scenarios (see Fig.~\ref{fig:0}).  We account for the secular hyperfine interaction along with the electron- and nuclear-Zeeman terms due to an inhomogeneous magnetic-field gradient. This inhomogeneity can lead to dynamics in the nuclear-spin system and a consequent decay of electron-spin coherence, similar to the case of spectral diffusion from nuclear dipolar interactions. We further show that a magnetic-field gradient can result in the failure of two typical techniques used to mitigate decoherence due to the nuclear environment: nuclear spin state narrowing and Hahn echo. State narrowing involves the preparation of the nuclear-spin bath in a narrow distribution of eigenstates of the nuclear field operator.  Ideally, this corresponds to an exact eigenstate of the component of the nuclear field along an applied magnetic field.\cite{Coish2004,PhysRevB.73.205302,Stepanenko2006enhancement,Giedke2006quantum,PhysRevLett.105.216803}  The presence of a transverse magnetic-field gradient destroys this dephasing-free nuclear-spin state, leading to free-induction decay (FID) of the spin qubit. The second technique, a Hahn echo (HE), involves a refocusing pulse applied at time $t$ after initial preparation. This procedure reverses the evolution of the spin qubit under a \emph{static} secular hyperfine coupling, allowing for a retrieval of coherence at time $2t$. In this case, the presence of an inhomogeneous transverse magnetic field then leads to a finite nuclear-spin-bath correlation time, preventing the full recovery of spin-qubit coherence. 

Our analysis of a transverse magnetic-field gradient on the decoherence dynamics of spin qubits applies to several different systems: spins in quantum dots formed in, e.g., either III-V semiconductors or Si, and to single P donor impurities in Si (Si:P) (see Fig.~\ref{fig:0}). In each case, we find that a magnetic-field gradient can decrease the coherence time, relative to known coherence times in the absence of a magnetic-field gradient. In addition to electron spins, our dephasing model is directly applicable to hole spins in III-V semiconductors, \cite{fischer2008spin, Eble2009hole, Brunner2009coherent, DeGreve2011ultrafast, PhysRevLett.109.237601} and may give insight into dephasing from the nuclear quadrupolar interaction due to inhomogeneous strain in InAs nanowires.\cite{nadj2010spin}

The remainder of this article is divided as follows. In Sec.~\ref{sec:model}, we introduce the Hamiltonian accounting for Zeeman terms coupling to the spin qubit and nuclear-spin bath for an inhomogeneous magnetic field, as well as the secular hyperfine coupling. We then derive an exact formula for the spin coherence factor. In Sec.~\ref{sec:analytic}, we introduce a perturbative method to obtain simple expressions for the coherence factor with associated parametric dependences. In Sec.~\ref{sec:spec}, we obtain analytical expressions for the electron spin coherence factor in the context of free-induction decay for an initial narrowed nuclear-spin state and for Hahn echo with an initial infinite-temperature thermal state of the nuclear-spin system. In Sec.~\ref{sec:app}, we apply our model to the physical systems shown in Fig.~\ref{fig:0} and make testable predictions for experiments that we expect to be realizable with present technology.  We conclude in Sec.~\ref{sec:Conclusions} with a discussion of the most important results presented here and possible implications for future experiments on spin dynamics in these systems.

\section{Hamiltonian and exact solution \label{sec:model}}

\begin{figure}
	\begin{center}
		\includegraphics[scale=1.2]{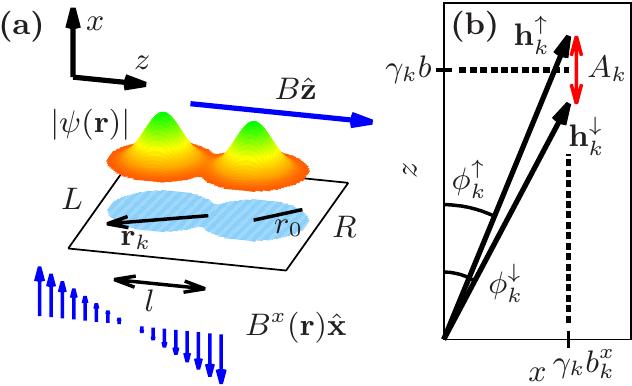}
	\end{center}
	\caption{(Color online) (a) Example of a device studied in this paper. A constant longitudinal magnetic field $B\hat{\mvec z}$ and an inhomogeneous transverse magnetic field $B^x(\mvec r)\hat{\mvec x}$ are applied to an electron spin interacting with a nuclear-spin bath. The light blue circles illustrate the range $r_0$ of the electron wavefunction $\psi(\mvec r)$ in the specific case of a double quantum dot with left ($\ket L$) and right ($\ket R$) single-particle orbital states separated by a distance $l$. (b) Total effective field $\mvec h_k^{\upa/\dwna}$ experienced by nuclear-spin $k$, including Knight shifts coming from the secular hyperfine interaction (of strength $A_k$) with the electron spin, which can be in state $\ket{\upa}$ or $\ket{\dwna}$. This can be viewed as an electron-spin-dependent rotation of the nuclear spin precession axis by an angle $\phi_k^{\upa/\dwna}$, which also depends on the Zeeman field $\gamma_k (b_k^x,0,b)$. Thus, because of the inhomogeneous magnetic field, the precession axis is slightly different for each nuclear spin.
 This leads to a finite correlation time for collective states of the nuclear-spin bath, preventing the full recovery of electron spin coherence. \label{fig:1}}
\end{figure}

We consider a single electron spin interacting with a nuclear-spin bath through hyperfine coupling.  In general, the coupling may arise from the Fermi contact term\cite{Fermi:1930fk} (most relevant for electrons in III-V semiconductors), purely anisotropic interactions (relevant for, e.g., hole spins in III-V semiconductors\cite{fischer2008spin}), or both (relevant for, e.g., spins in silicon\cite{Witzel2007,saikin2003nonideality}). We will, however, consider only the cases where spin dynamics is dominated by the secular hyperfine coupling (the part that commutes with the electron Zeeman term).  We allow for a position-dependent external magnetic field $\mvec B_\mrm{ext}(\mvec r)=B^x(\mvec r)\,\mvec{\hat x}+B\,\mvec{\hat z}$, $B$ being a constant, as illustrated in Fig.~\ref{fig:1}(a). We further take $\mvec B_\mrm{ext}(\mvec r)$ and the electron envelope wavefunction $\psi(\mvec r)$ to satisfy $\int d^3 r \mvec B_\mrm{ext}(\mvec r)|\psi(\mvec r)|^2 =B\,\mvec{\hat z}$ [this is true, e.g., when $B^x(\mvec r)$ is an odd function of $z$, while $|\psi(\mvec r)|$ is an even function of $z$, as in Fig.~\ref{fig:1}(a)]. As illustrated in Fig.~\ref{fig:1}(a), this means that the external magnetic field experienced by the electron is aligned with the quantum dot principal axis of symmetry, such that the $g$-tensor is diagonal and only its $z$ component is relevant. Adding a secular hyperfine coupling gives\footnote{Anisotropic hyperfine interactions of the form $\sum_k (A_k^x \Ikx+A_k^z\Ikz)\hat S_z$ can easily be accounted for within our theoretical framework if each nuclear spin is rotated by an angle $\theta_k=\arctan(A_k^x/A_k^z)$ around the $y$-axis by applying the operators $\hat R_k^y(\theta)=\exp(-i\theta_k\Iky)$. We retrieve a Hamiltonian that has the exact same form as Eq.~(1), but with modified coupling constants $A_k\rightarrow\sqrt{(A_k^x)^2+(A_k^z)^2}$, gyromagnetic ratios $\gamma_k\rightarrow\gamma_k(b_k^xA_k^x+b A_k^z)/bA_k'$, and transverse fields $b_k^x\rightarrow b(b_k^xA_k^z-bA_k^x)/(bA_k^z+b_k^x A_k^x)$. However, this approach will not reproduce exactly the same Hamiltonian in the presence of terms proportional to $\sum_k A_k \Iky\hat S^z$, but should lead to similar qualitative behavior, especially if the anisotropic hyperfine coupling terms are weak.} (setting $\hbar=1$),

\be
	\hat H=(b+\hz) \Sz+\sum_k\gamma_k\left(b_k^x\Ikx+b\Ikz\right),	\label{eqn:H}
\ee
where $\hat h^z=\sum_kA_k\Ikz$ is the Overhauser field, with sums over nuclear spins $k$, $b=g^\ast\mu_B B$ is the electron Zeeman splitting, $\gamma_k b_k^x=\gamma_k g^\ast\mu_B B^x(\mvec r_k)$, and $\gamma_k=\gamma_k^\mrm I/g^\ast\mu_B$, where $\gamma_k^\mrm I$ is the nuclear gyromagnetic ratio. The nuclear gyromagnetic ratios are thus given by $\gamma_k g^\ast\mu_B$, with $\mu_B$ the Bohr magneton and $g^*$ the effective $g$-factor. Finally, $A_k \propto |\psi(\mvec r_k)|^2$ is the hyperfine coupling strength at nuclear site $k=0,1,2,\ldots$ These couplings are calculated as functions of $k$ for several relevant wavefunctions in Appendix~\ref{sec:A}.

Eigenstates of $\hat H$, Eq.~\eq{eqn:H}, are simultaneous eigenstates of $\Sz$, which we denote $\{\kup,\kdwn\}$.
Since there is no interaction between nuclear spins, the eigenstates of $\hat H$ can further be written as a product of nuclear-spin states of definite angular momentum along the direction of an effective field $\mathbf{h}_k^\sigma$, [see Fig. \ref{fig:1}(b)], for each nuclear spin at site $k$ with electron spin $\sigma=\uparrow,\downarrow$.  These states can be obtained from an electron-spin-state-dependent rotation $\hat R_k^{\upa/\dwna}=\eul{-i\Iky\phi_k^{\upa/\dwna}}$ applied to $I_k^z$-eigenstates, $\left|I_k,m_k\right>$, where
\be
		\phi_k^{\upa/\dwna}=\arctan\left(\frac{\gamma_kb_k^x}{b\gamma_k\pm \frac12A_k}\right),
\ee
as illustrated in Fig.~\ref{fig:1}(b). This gives the exact electron-spin coherence factor $C(t)$, defined by $\mean{\hat S^+(t)}=\mean{\hat S^+(0)}C(t)$, for two protocols: (i) free-induction decay and (ii) Hahn echo.

For both Hahn echo and free-induction decay, we take an initial nuclear-spin state of the form $\hat \rho_\mrm I (0)=\sum_{j=1}^M p_j\proj j$, with $\ket j=\prod_k\ket{\psi_k^j}$, i.e., a statistical mixture containing $M$ tensor-product states. In this paper, we will focus on two possible initial states: narrowed states and infinite-temperature thermal states. Narrowed nuclear-spin states are defined as a mixture of eigenstates $\{\ket{j}\}$ of $\hat{h}^z$ having the same eigenvalue $h^z_n$ for all $j$, i.e. that satisfy the equation
\begin{equation}
 \hat h^z\ket j=h^z_n\ket j.\label{eqn:def:narrowed}
\end{equation}
For numerical evaluation, we generate such a state by taking $\ket{\psi_k^j}=\ket{I_k,m_k^j}$ and $p_j = \delta_{jj'}$ (i.e., $M=1$), with the eigenvalues $m_k^j$ of $\Ikz$ uncorrelated for different $k$ and chosen randomly from a uniform distribution. To evaluate thermal averages over infinite-temperature states, we take $p_j=1/M$ with $\ket{\psi_k^j}$ a random spin coherent state with quantization axis sampled uniformly on the unit sphere independently for each $k,j$.  The value of $M$ is set by increasing this value until the coherence factor has converged to within approximately $1\%$; we find a value $M\sim 100$ is typically sufficient for convergence.


In the case of free-induction decay, the time-evolution operator is simply $\hat U(t)=\eul{-i\hat Ht}$ and the coherence factor is given by $C(t)=\sum_jp_jC_j(t)$, with 
\be
	C_j(t)=\prod_k\bra{\psi_k^j}\hat R_k^{\upa}[\hat E_k^\upa(t)]^\dagger \hat Q_k^\dwna \hat E_k^\dwna(t)(\hat R_k^\dwna)^\dagger\ket{\psi_k^j},\label{eqn:C:FID}
\ee
and where we have introduced $\hat Q_k^{\ud}=\eul{-i\Iky(\phi_k^{\upa/\dwna}-\phi_k^{\dwna/\upa})}$ and
\begin{align}
	\hat E_k^\ud(t)&=\sum_{m_k=-I_k}^{I_k}\eul{-im_kh_k^\ud t}\proj{I_k,m_k},\\
	h_k^{\upa/\dwna}&=\sqrt{\left(\gamma_kb_k^x\right)^2+\left(b\gamma_k\pm A_k/2\right)^2}.
\end{align}
In the case of Hahn echo, a refocusing pulse (a $\pi$-pulse, with unitary given by the Pauli matrix $\sx$) is applied at time $t$, and again at time $2t$ to return the spin to its initial state. Thus, the time-evolution operator is given by $\hat U(2t)=\eul{-i\sx\hat H\sx t}\eul{-i\hat H t}$ and the electron-spin coherence factor is
\be
	C_j(2t)=\prod_k\bra{\psi_k^j}\hat R_k^\upa(\hat E_k^\upa)^\dagger \hat Q_k^\dwna (\hat E_k^\dwna)^\dagger \hat Q_k^\upa \hat E_k^\upa \hat Q_k^\dwna \hat E_k^\dwna(\hat R_k^\dwna)^\dagger\ket{\psi_k^j}.\label{eqn:C:HE}
\ee
Note that we have dropped the explicit $t$-dependence on $\hat{E}_k^\ud$ in Eq. \eqref{eqn:C:HE}.

Knowing the distribution of hyperfine couplings, magnetic-field distribution, and gyromagnetic ratios for the nuclear spins, Eqs.~\eq{eqn:C:FID} and~\eq{eqn:C:HE} can be used to compute the exact electron-spin coherence factor numerically. The calculation involves a product of $\mathcal O(N)$ square matrices of dimension $2I+1$, $N$ being the number of nuclear spins within $r_0$, the effective (donor or quantum-dot) Bohr radius. Therefore, for large systems with $\gtrsim10^6$ nuclear spins, the computation time can become rather long. More importantly, Eqs.~\eq{eqn:C:FID} and~\eq{eqn:C:HE} do not give any physical insight into the coherence time of the electron spin with respect to relevant physical parameters. For these two reasons, in the rest of this paper, we seek approximate expressions for the coherence factor which have a simpler form.

\section{Simplified coherence factor\label{sec:analytic}}

In this section, we approximate $C(t)$ using a Magnus expansion. The Magnus expansion is a perturbation theory in the amplitude of Knight-shift fluctuations $\sim A_k$ relative to the typical rate of nuclear-spin fluctuations $\sim \omega_k=\gamma_k\sqrt{b^2+(b_k^x)^2}$.  We will be able to truncate the expansion for the time-evolution operator at leading order in $A_k/\omega_k \ll 1$, giving expressions for both free-induction decay and Hahn echo. In addition, we will invoke a Gaussian approximation, valid for a large uncorrelated nuclear-spin bath, to obtain simple approximate expressions for the coherence factor. We ultimately show that, at leading order in the Magnus expansion, the coherence factor for a nuclear-spin bath in a narrowed state is only affected by fluctuations of transverse components of the nuclear field, $\hat h_x,\hat h_y$, whereas all components ($\hat h_x, \hat h_y, \hat h_z$) are equally important for an infinite-temperature thermal state. This approach generalizes that applied in the recent work of Ref.~\onlinecite{PhysRevLett.109.237601} on hole-spin dynamics.
\vspace{1cm}

\subsection{Time-evolution operators\label{sec:evolution}}

We first move the Hamiltonian of Eq.~\eq{eqn:H}, $\hat H = \hat H_0 + \hat V$, to the interaction picture with perturbation $\hat V=\hz \Sz$ and taking $\hat H_0$ to be the Zeeman terms, yielding
\begin{align}
	\hat V(t)&=\eul{i\hat H_0t}\hat V\eul{-i\hat H_0t}=\Sz\sum_kA_k\mvec v_k(t)\cdot\vecI_k,\\
	\mvec v_k(t)&=\left[-n_k^x(\cos\omega_kt-1),n_k^x\sin\omega_kt,(n_k^z)^2+(n_k^x)^2\cos\omega_kt\right],\label{eqn:vk}
\end{align}
where $\mvec n_k=\gamma_k(b_k^x,0,b)/\omega_k$.

In the Magnus expansion,\cite{burum1981magnus,PhysRevB.25.6622} the time-evolution operator is recast in the form $\hat U(t)=\eul{-i \hat H_\mrm M (t)}$, where $\hat H_\mrm M(t)$ is given by a series expansion, $\hat H_\mrm M(t)=\sum_{n=0}^\infty \hat H^{(n)}(t)$. The term $\hat H^{(n)}(t)$ contains $n+1$ time integrals over the rapidly oscillating perturbation $\hat V(t)$. We therefore expect rapid convergence of the Magnus expansion in the limit of large $\omega_k$, which sets the oscillation frequency for $\hat V(t)$. Explicit formulas for the lowest orders of the Magnus expansion are given in the literature.\cite{burum1981magnus,PhysRevB.25.6622} Criteria for the convergence of the Magnus expansion are discussed for specific physical systems considered here in Appendix~\ref{sec:Magnus}.

At leading order in the Magnus expansion, $\hat U(t)\simeq\eul{-i\hat H^{(0)}(t)}$. As will be shown below, this leading-order analysis is sufficient to describe the dynamics of the coherence factor in several different spin systems. In the case of free-induction decay,
\begin{widetext}
\begin{align}
	\hat H^{(0)}(t)&=\int_0^t\hat V(t_1)dt_1=\Sz\sum_k\mvec h_k^\mrm{FID}(t)\cdot\hat{\mvec I}_k,\label{eqn:def:h:FID}\\
	\mvec{h}_k^\mrm{FID}(t)&=A_k\left[n_k^xn_k^z\left(t-\frac{\sin\omega_kt}{\omega_k}\right),n_k^x\frac{1-\cos\omega_kt}{\omega_k},(n_k^z)^2t+(n_k^x)^2\frac{\sin\omega_kt}{\omega_k}\right].\label{eqn:h:FID}
\end{align}
For Hahn echo, $\pi$-pulses are applied at times $t$ and $2t$, leading to $\Sz\rightarrow\sx\Sz\sx=-\Sz$. Therefore,
\begin{align}
	\hat H^{(0)}(2t)&=\Sz\sum_kA_k\vecI_k\cdot\left[\int_0^t\mvec v_k(t_1)dt_1-\int_t^{2t}\mvec v_k(t_1)dt_1\right]=\Sz\sum_k\mvec{h}_k^\mrm{HE}(t)\cdot\vecI_k,\label{eqn:def:h:HE}\\
	\mvec{h}_k^\mrm{HE}(t)&=\frac{A_k b_k^x}{\omega_k^2}\left[n_k^z f_k^z(t),-f_k^y(t),-n_k^xf_k^z(t)\right],\label{eqn:h:HE}\\
	\mbox{where }f_k^y(t)&=2\cos\omega_kt-\cos2\omega_kt-1,
	\qquad f_k^z(t)=\sin2\omega_kt-2\sin\omega_kt.
\end{align}
\end{widetext}
In both cases (free-induction decay and Hahn echo), we have found an approximate evolution operator of the form $\exp[-i\Sz\hat X(t)]$, where $\hat X(t)=\sum_k\mvec h_k(t)\cdot\vecI_k$ has the simple form of independent time-varying effective fields $\mvec h_k(t)$ on each nuclear spin. As expected, if $b_k^x=0\;\forall\;k$, $\mvec h_k^\mrm{FID}(t)$ fluctuates only along $\hat{z}$, resulting in pure dephasing on a time scale $T_2^\ast$ for a random initial nuclear-spin state. This decay will not occur, however, if the nuclear bath is initially in a narrowed state. For the ideally narrowed initial state defined by Eq.~\eq{eqn:def:narrowed}, dephasing will arise entirely from the transverse components, $h_k^{x,\mrm{FID}}(t), h_k^{y,\mrm{FID}}(t)$, in the presence of finite $b_k^x$. Finally, for Hahn echo, the evolution operator only deviates from the identity for non-zero values of $b_k^x$.

\subsection{Gaussian approximation and finite-size effects\label{sec:gauss}}

The steps taken above have allowed us to obtain a simpler approximate evolution operator in the interaction picture. Nevertheless, applying this simplified evolution operator to find the coherence factor $C(t)$ typically leads to complicated expressions, especially for the case of large nuclear spin, $I>1/2$. Therefore, we will take advantage of the large number of nuclear spins usually present in semiconductor devices to introduce the Gaussian approximation.

We will use the same symbol, $\hat H^{(0)}(\tau)$, for the leading-order term in the Magnus expansion for both Hahn echo and free-induction decay.  For the case of free-induction decay, we take $\tau=t$  and for Hahn echo, $\tau=2t$. The transverse spin is then given by 
\be
	\mean{\Spl(\tau)}=\eul{i\phi(\tau)}\mean{\eul{i\mrm L_0(\tau)}\Spl},
\ee
where $\phi(\tau)=bt$ for free-induction decay and $\phi(\tau)=0$ for Hahn-echo. We have also introduced $\mrm L_0(t)\hat O=[\hat H^{(0)}(\tau),\hat O]$, i.e., $\mrm L_0(t)$ is the Liouvillian superoperator associated with $\hat H^{(0)} (\tau)$. In both Eqs.~\eq{eqn:def:h:FID} and~\eq{eqn:def:h:HE}, $\hat H^{(0)}(\tau)$ has the form $\hat H^{(0)}(\tau)=\Sz\hat X(\tau)$, where $\hat X(\tau)=\sum_{k}\mvec h_k(\tau)\cdot\hat {\mvec  I}_k$ acts only on nuclear-spin degrees of freedom.  This important property allows us to calculate each power in the Taylor series expansion of $\eul{i\mrm L_0(\tau)}\Spl$, giving $\eul{i\mrm L_0(\tau)}\Spl=\exp[i \hat X (\tau)]\hat S^+$. For an initial state of the form $\hat\rho_\mrm S(0)\otimes\hat\rho_\mrm I (0)$, with $\hat\rho_\mrm S(0)$ and $\hat\rho_\mrm I(0)$ respectively the initial spin-qubit and nuclear-spin states, we find $\mean{\Spl(\tau)}=\mean{\hat S^+(0)}C_\chi(\tau)$. This defines the coherence factor for a nuclear-spin bath initially in state $\rho_I(0)= \chi$: $C_\chi(\tau)=\eul{i\phi(\tau)}\mean{\exp[i\hat X(\tau)]}_\chi$. Since dephasing arises from fluctuations in the nuclear field, we define $\delta_\chi\hat  X(\tau)=\hat X(\tau)-\mean{\hat X(\tau)}_\chi=\sum_k \mvec h_k(\tau)\cdot\delta_\chi\hat{\mvec I}_k$, and the coherence factor becomes
\be
	C_\chi(\tau)=\eul{i\phi(\tau)}\eul{i\mean{\hat X(\tau)}_\chi}\bmean{\eul{i\delta_\chi \hat X(\tau)}}_\chi.\label{eqn:exp:exact}
\ee
The Gaussian approximation then corresponds to taking $\mean{\exp[i\delta_\chi \hat X(\tau)]}_\chi\simeq\exp[-\frac12\mean{[\delta_\chi\hat X(\tau)]^2}_\chi]$. This approximation is justified when the following two conditions are met:
\begin{enumerate}
	\item There are no correlations between nuclear spins, i.e. $\mean{\delta_\chi I_{k_1}^{l_1}\delta_\chi I_{k_2}^{l_2}}_\chi=\mean{\delta_\chi I_{k_1}^{l_1}}_\chi\mean{\delta_\chi I_{k_2}^{l_2}}_\chi\;\forall\;k_1\neq k_2$, with $l_1,l_2\in\{x,y,z\}$.
	\item $N\gg 1$, where $N$ is the number of nuclear spins within the range of the electron wavefunction.
\end{enumerate}
The first condition is always met for the infinite-temperature states introduced in Section~\ref{sec:model}. As will be seen later in this Section, it can also be satisfied for the ideally narrowed state defined by Eq.~\eq{eqn:def:narrowed}. However, the number of nuclear spins $N$ needed to adequately satisfy the second criterion can be very large. Indeed, as discussed in Appendix~\ref{sec:Gaussian}, subleading corrections to the Gaussian approximation are typically only suppressed by $\sim \mathcal{O}\left(1/N^\alpha\right)$, with $\alpha<1$ ($\alpha=1/4$ for free-induction decay and $\alpha=1/8$ for Hahn echo).  Thus, non-Gaussian corrections can have an important effect in small systems such as single phosphorus donors in silicon, where $N\sim 10^2$. As will be shown in Section~\ref{sec:Si:P}, in situations where the Gaussian approximation predicts Gaussian ($\sim\eul{-t^2}$) decay, the exact solution will rather exhibit exponential ($\sim\eul{-t}$) behavior.

For the rest of this section, we assume that $N$ is sufficiently large to avoid these finite-size effects. We can then proceed to the calculation of explicit coherence-factor formulas for each considered nuclear-spin state. We first consider an infinite-temperature thermal state, which we model as described in Section~\ref{sec:model}, i.e. by a statistical mixture $\hat\rho_\mrm I(0)=\sum_jp_j\proj{j}$ of $M$ states $\ket j=\prod_k\ket{\psi_k^j}$, with each nuclear spin randomly oriented in space. In this state, nuclear spins are uncorrelated, justifying use of the Gaussian approximation. Calculating the moments involved in Eq.~\eq{eqn:exp:exact} and taking $M\rightarrow\infty$, we find that the only nonvanishing contributions come from $\mean{(\Ikx)^2}_\mrm{th}=\mean{(\Iky)^2}_\mrm{th}=\mean{(\Ikz)^2}_\mrm{th}=\frac13I_k(I_k+1)$, and the coherence factor reduces to
\begin{align}
	C_\mrm{th}(\tau)&\simeq\eul{i\phi(\tau)}\exp\left[-\frac16\sum_kI_k(I_k+1)|\mvec h_k(\tau)|^2\right].\label{eqn:C:th}
\end{align}
Thus, for an infinite-temperature thermal state, all components of the nuclear field ($\hat h_k^x, \hat h_k^y, \hat h_k^z$)  appear with the same prefactor, with each individual value of $h_k^\alpha(\tau)$ set by the field distribution and coherence measurement scheme (free-induction decay or Hahn echo). 

By definition, an ideally narrowed state $\hat\rho_n$ satisfies Eq.~\eq{eqn:def:narrowed}. Most generally, this state can be written as $\hat \rho_n=\sum_j p_{j}\proj j+\sum_{i\neq j}\rho_{ij}\ket i\bra j$, where the sums are performed over all the degenerate eigenstates $\ket j$ of $\hat h^z$ with energy $h^z_n$. We can then use the fact that $C_n(\tau)=\eul{i\phi(\tau)}\mean{\exp[i\hat X(\tau)]}_n=\eul{i\phi(\tau)}\sum_j p_j C_j(\tau)+\eul{i\phi(\tau)}\sum_{i\neq j} \rho_{ij} C_{ij}(\tau)$, where $C_j(\tau)=\mean{\exp[i\hat X(\tau)]}_j$ and $C_{ij}(\tau)=\bra j\exp[i\hat X(\tau)]\ket i$. Assuming no special phase relationship between the eigenstates $\ket j$, we can drop the second term in $C_n(\tau)$, which amounts to dropping the coherences in $\hat\rho_n$.\cite{PhysRevB.81.165315} Then, for each tensor-product state $\ket j=\prod_k\ket{I_k,m_k^j}$, nuclear-spin fluctuations on different sites are uncorrelated (i.e., $\mean{\delta_j \hat I_k^z\delta_j \hat I_{k'}^z}_j\propto \delta_{k,k'}$), so we apply the Gaussian approximation to calculate each $C_j(\tau)$. We further find that all the moments involved vanish in each state $\ket j$, except for $\mean{\hat I_k^z}_j$,  $\mean{(\delta_j\hat I_k^x)^2}_j$, $\mean{(\delta_j\hat I_k^y)^2}_j$, $\mean{\delta_j \Ikx\delta_j\Iky}_j$, and $\mean{\delta_j \Iky\delta_j\Ikx}_j$. Furthermore, we find that $\mean{\delta_j \Ikx\delta_j\Iky}_j$ and $\mean{\delta_j \Iky\delta_j\Ikx}_j$ cancel each other out, and that $\mean{\Ikz}_j=m_k^j$ and $\mean{(\delta_jI_k^x)^2}_j=\mean{(\delta_jI_k^y)^2}_j=\frac12[I_k(I_k+1)-(m_k^j)^2]$. The result is further simplified for $N\gg1$. Indeed, the number of available eigenstates $\ket j$ of $\hat h^z$ scales exponentially with $N$, such that states that have isotropic distributions of $m_k^j$'s are overwhelmingly more probable than states for which $m_k^j$ depends on the site $k$. We can thus replace every $(m_k^j)^2$ by its expectation value\footnote{
	Deviations from the predictions of the approximation $(m_k^j)^2\rightarrow \sum_m p(m) m^2$ were numerically checked to be negligible even for systems as small as single phosphorus donors in silicon, where $N\simeq 250$. This is not surprising, since this approximation works when there is a large number of nuclear-spin configurations that lead to the same average polarization in a small region of space. That number of configurations, and thus the quality of the approximation, grows exponentially with $N$, in contrast with the Gaussian approximation, which improves as a power law in $N$.
} $E(m^2)=\sum_{m}p(m)m^2$, with $p(m)$ a probability distribution over accessible values of $m$. Taking $p(m)$ to be uniform, we find $\mean{(\delta_jI_k^x)^2}_j=\mean{(\delta_jI_k^y)^2}_j=\frac13I_k(I_k+1)$. Inserting the calculated moments in Eq.~\eq{eqn:exp:exact} and considering a single realization $j$ of a narrowed state, we obtain
\be
	C_\mrm{n}(\tau)\simeq\eul{i\phi_\mrm n(\tau)}\exp\left[-\frac16\sum_kI_k(I_k+1)|\mvec h_k^\bot(\tau)|^2\right],\label{eqn:C:n}
\ee
where $\mvec h_k^\bot(\tau)$ is the projection of $\mvec h_k(\tau)$ in the $x-y$ plane and $\phi_\mrm n(\tau)=\phi(\tau)+\frac{h^n_z} A\sum_kh_k^z(\tau)$, with $h_z^n=\bra{j}\hat{h}^z\ket{j}=\sum_k A_k m_k^j$. Strikingly, $h_k^z(\tau)$ appears only in the phase $\phi_n(\tau)$, and therefore does not contribute to the decay of the coherence factor. This is a direct consequence of the vanishing variance in $\hat h_z$ for a narrowed state.

\section{Coherence measurement protocols\label{sec:spec}}

In the previous section, we found compact expressions for $C(\tau)$ for a nuclear-spin bath initially in an infinite-temperature thermal or narrowed state. We now use these results to obtain characteristic features of the coherence dynamics under a magnetic-field gradient. In other words, we replace $\mvec h_k(\tau)  \to \mvec h_k^\mathrm{FID}(\tau)$ for free-induction decay and $\mvec h_k(\tau)  \to \mvec h_k^\mathrm{HE}(\tau)$ for Hahn echo. In each case, we study both short-time and long-time behavior. We first consider free-induction decay for an initially narrowed nuclear-spin state and show that at long time ($t\gtrsim \max[1/\gamma\Delta b^x,1/\gamma b]$, with $\Delta b^x$ the typical range of $b^x_k$ experienced within the envelope wavefunction), the coherence factor decays as a Gaussian, $C(t)\sim \exp[-(t/T_2^\nabla)^2]$. We then address the case of Hahn-echo decay with infinite-temperature thermal states. Although for short times, we find the Hahn-echo decay envelope to be of the form $\sim\exp[-(t/T_{2e}^\nabla)^4]$, for long times coherence dynamics is very rich, displaying revivals, incomplete decay due to motional averaging, or exponential decay (reflecting a Markovian limit), depending on the particular physical setting and associated parameters.

\subsection{Free-induction decay with narrowed states\label{sec:narrowed:FID}}

Substituting Eq.~\eq{eqn:h:FID} for $\mvec h_k(\tau)$ into Eq.~\eq{eqn:C:n}, which gives the coherence factor for a bath initially in a narrowed state, we obtain
\begin{align}
	C_\mathrm{n}(t)&\simeq\eul{i\phi_\mrm n(t)}\exp\left[\!-\!\sum_k\frac{I_k(I_k+1)}{6}A_k^2(n_k^x)^2g_k(t)\right]\!\!,\label{eqn:C:int:FID:n}\\
	g_k(t)&=(n_k^z)^2\left(t-\frac{\sin\omega_kt}{\omega_k}\right)^2+\left(\frac{1-\cos\omega_kt}{\omega_k}\right)^2.\label{eqn:g}
\end{align}
The term that grows as $\sim t^2$ in the argument of the exponential in $C(t)$ dominates for  $t\gtrsim\max(1/\gamma \Delta b^x,1/\gamma b)$, resulting in an approximate Gaussian decay,
\begin{align}
	C_\mathrm{n}(t)&\simeq\eul{i\phi_\mrm n(t)}\exp\left[-\left(t/T_2^\nabla\right)^2\right],\\
	\frac{1}{T_2^\nabla}&=\frac1b\sqrt{\frac16\sum_s\nu_sI_s(I_s+1)\gamma_s^2\Sigma_s^2},\label{eqn:T2nabla}\\
	\Sigma_s^2&=\sum_{k_s}(b_{k_s}^xA_{k_s})^2,\label{eqn:def:sigma2}
\end{align}
where $\sum_s$ is taken over all nuclear species $s$ in the material, such that $\nu_s$, $I_s$, and $\gamma_s$ are, respectively, the relative abundance, nuclear spin, and gyromagnetic ratio of species $s$. We have also defined $\Sigma_s^2$ as a sum over nuclei $k_s$ belonging to the same species $s$, which depends on the geometry of the magnetic field through $b^x_{k_s}$ and on the electron wavefunction through $A_{k_s}$. Explicit formulas for $\Sigma_s^2$ are given in Appendix~\ref{sec:Sigma2} for various geometries. Thus, in this typical narrowed-state free-induction decay scenario, we find Gaussian decay with inverse decay time $1/T_2^\nabla\sim (A/\sqrt N)(\gamma\Delta b^x/b)$. As mentioned in Section~\ref{sec:gauss} and as will be shown explicitly in Section~\ref{sec:Si:P}, this behavior results from the Gaussian approximation, which breaks down when the number $N$ of nuclear spins interacting with the electron spin is not sufficiently large.

\subsection{Hahn echo\label{sec:thermal:HE}}

We now consider the coherence factor when the nuclear-spin bath is in an infinite-temperature thermal state.  The relevant coherence factor is given by Eq.~\eq{eqn:C:th}. Substituting $\mvec h_k(t)\to\mvec h^\mathrm{HE}_k(t)$ from Eq.~\eq{eqn:h:HE} into Eq.~\eq{eqn:C:th}, we find
\begin{align}
	C_\mathrm{th}(2t)&\simeq\exp\left[-\frac{8}{3}\sum_kI_k(I_k+1)\frac{(A_{k}\gamma_kb_{k}^x)^2}{\omega_{k}^4}\sin^4\left(\frac{\omega_{k}t}{2}\right)\right].\label{eqn:C:HE:th}
\end{align}
This sum can be simplified further in various physically meaningful cases which are discussed in this section.

The most straightforward way to reduce Eq.~\eq{eqn:C:HE:th} is through a short-time approximation. Expanding $\sin^4(\omega_k t/2)$ to leading order in $\omega_kt/2$ yields
\begin{align}
	C_\mrm{sh}(2t)&\simeq\exp\left[-\left(t/T_{2,e}^{\nabla}\right)^4\right],\label{eqn:sh}\\
	\frac{1}{T_{2,e}^\nabla}&=\left(\frac16\sum_s\nu_sI_s(I_s+1)\gamma_s^2\Sigma_s^2\right)^{1/4}.\label{eqn:T2e}
\end{align}
Thus, under the short-time approximation, decay occurs within a dephasing time $T_{2,e}^\nabla$, defined such that $C_\mrm{sh}(2T_{2,e}^\nabla)=1/\mrm{e}$. As will be shown in Section~\ref{sec:GaAs}, this dephasing mechanism can dominate other processes (electron-nuclear flip-flops and nuclear dipolar interactions), which have given rise to decay measured in GaAs singlet-triplet qubits.\cite{bluhm2010dephasing,PhysRevB.77.125329,PhysRevB.74.035322} In addition, we further note that a gradient in the $z$ component of the magnetic field would have no effect on Hahn-echo dephasing. Indeed, such a gradient can easily be incorporated in our model with the simple replacement $\gamma_k b\rightarrow\gamma_k b_k^z$. Since $T_{2,e}^\nabla$ is independent of $b$, $z$ gradients do not contribute; in order for a longitudinal gradient to contribute to dephasing, nonsecular terms in the hyperfine interaction would need to be included.\cite{hung2013hyperfine}

As in recent calculations for hole-spin echo dynamics,\cite{PhysRevLett.109.237601} for large $b$, a motional-averaging regime can be reached in Hahn echo. Indeed, when $b\gg\Delta b^x$, we have $\omega_{k_s}\simeq \gamma_s b$ and then Eq.~\eq{eqn:C:HE:th} predicts recurrences in the coherence factor with a period $\sim 1/\gamma_s b$.  For sufficiently large $\omega_k\propto b$, Eq.~\eq{eqn:C:HE:th} shows that $C(t)\sim 1$ for all time, in which case the motional-averaging regime has been reached. This occurs for $b\gtrsim b_c$, where
\be
	b_c=\left(\sum_s\nu_sI_s(I_s+1)\frac{\Sigma_s^2}{\gamma_s^4}\right)^{1/4}.\label{eqn:Bc}
\ee
Thus, while a transverse magnetic-field gradient can enhance dephasing, this enhancement can be controlled or eliminated with $b$ large enough to reach the motional-averaging regime. In GaAs, for a single electron shared by two dots of radius $r_0=25$~nm separated by $l=200$~nm with $N=4.4\times10^6$ nuclei within $r_0$ (parameters taken from Ref. \onlinecite{bluhm2010dephasing}) with an added transverse gradient $\left.\partial_zB^x\right|_{z=0}=1$~T$/\upmu$m (a typical value from Refs.~\onlinecite{pioro2008electrically,PhysRevB.81.085317,PhysRevLett.107.146801}), we find $B_c=b_c/g^\ast\mu_B\sim300$~mT. For a single dot with the same properties, we find $B_c\sim200$~mT. To evaluate $\Sigma_s^2$, we have used Eqs.~\eq{eqn:sigma2:double:1e} and~\eq{eqn:sigma2:single}, respectively, for double- and single-dot geometries.

Finally, when $b\ll\Delta b^x$ we replace $\omega_k \simeq\gamma_kb_k^x$ in Eq.~\eq{eqn:C:HE:th}. The time-dependence of the resulting sum over $k$ then strongly depends on geometry. Therefore, we have derived explicit formulas for that sum limiting ourselves to single electrons in single and double dots in 2D, with Gaussian orbital wavefunctions. In both cases, we take $b_k^x$ from Eq.~\eq{eqn:bkx} and $A_k$, respectively, from Eq.~\eq{eqn:Ak:single} or~\eq{eqn:Ak:double:1e} for single and double dots. Assuming an isotropic distribution of nuclear spins, we calculate an average of the sum over $k$ appearing in Eq.~\eq{eqn:C:HE:th} with respect to the angular degree of freedom in the position of each nucleus. For $t\gtrsim 1/\gamma\Delta b^x$, for single dots the decay becomes exponential, with $C_\mrm{th}(2t)=\exp(-t/T_\mrm{2M}^\nabla)$. For a Gaussian electron wavefunction, the $A_k$'s are given by Eq.~\eq{eqn:Ak:single} and we obtain for $b\ll\Delta b^x,t\gtrsim 1/\gamma\Delta b^x$,
\begin{eqnarray}
C_\mrm{th}(2t)&=&\exp(-t/T_\mrm{2M}^\nabla)\\
\frac{1}{T_\mrm{2M}^\nabla}&=&\frac{\sqrt\pi}{24}\sum_s\nu_s\frac{I_s(I_s+1)}{\gamma_s\delta b^x}\frac{A_s^2}{N},\label{eqn:T2:markov}
\end{eqnarray}
where $A_s$ is the total hyperfine coupling strength for nuclear spins of species $s$.  We have also defined $\delta b^x\equiv r_0\left.\partial_zb^x\right|_{z=0}$, the variation of $b_x$ over the single-dot Bohr radius $r_0$. This is to be distinguished from $\Delta b^x$, the variation of $b^x$ over the length-scale of the whole device, which becomes $\Delta b^x=l\left.\partial_zb^x\right|_{z=0}$ in the case of a double dot with separation $l$. Thus, Eq.~\eq{eqn:T2:markov} shows that the decay process becomes Markovian, leading to a pure exponential decay of $C_\mrm{th}(t)$, when the correlation time of the nuclear-spin bath $\sim 1/\gamma\Delta b^x$ is short compared to the decay time $T_\mrm{2M}^\nabla$, i.e., when $T_\mrm{2M}^\nabla>1/\gamma\Delta b^x$. This allows us to determine a critical field gradient, $\delta b^x_\mrm{M}$, beyond which the leading contribution to dephasing is exponential. Comparing Eqs.~\eq{eqn:sh} and~\eq{eqn:T2:markov} for a homonuclear (single-isotope) spin bath, we find
\be
	\delta b^x_\mrm M=\frac{1}{2\sqrt 3}\left(\frac\pi2\right)^{1/3}\sqrt{I(I+1)}\frac{A}{\gamma\sqrt N}.\label{eqn:deltabx:markov}
\ee
Thus, we find Markovian decay when the broadening of nuclear spin precession frequencies due to the gradient $\sim \gamma \delta b^x_\mrm M$ exceeds the nuclear-field fluctuations $\sim A/\sqrt N$. This behavior is not obtained in our model for double dots. Indeed, in that geometry, all nuclei that interact significantly with the electron spin are subject to a finite magnetic field, with average value $\pm (l/2)\left.\partial_zB^x\right|_{z=0}$, with $l$ the inter-dot spacing, the sign depending on which dot a nuclear-spin occupies. Averaging over the angular degree-of-freedom and evaluating the sum in Eq.~\eq{eqn:C:HE:th}, we find that the double-dot geometry prevents the occurrence of a Markovian regime at long times, and that we rather have $\lim_{t\rightarrow\infty}C_\mrm{th}(2t)=C_0$, $C_0$ being a constant. In other words, in a double quantum dot, a very strong transverse field gradient alone can lead to motional averaging, in contrast to the single-dot case where a longitudinal field is needed. This qualitative difference in behavior between single and double dots is illustrated in Fig.~\ref{fig:Si}(c).

\section{Physical Realizations\label{sec:app}}

In the previous sections, we have found the main features of the coherence dynamics under a magnetic-field gradient. The purpose of the present section is to predict whether these features could be measured in common materials used in current-day experiments. In particular, we find conditions under which this decoherence mechanism dominates over the leading dephasing sources known in these materials in the absence of a magnetic-field gradient. We focus on quantum dots in both GaAs and Si, and also investigate the case of single P donors in Si. Unless otherwise specified, for numerical evaluation, we have used hyperfine coupling constants and gyromagnetic ratios for the considered materials from the literature.\cite{PSSB:PSSB200945229,PhysRevB.83.165301}

\subsection{GaAs quantum dots\label{sec:GaAs}}

\begin{figure}
	\begin{center}
		\includegraphics[scale=1]{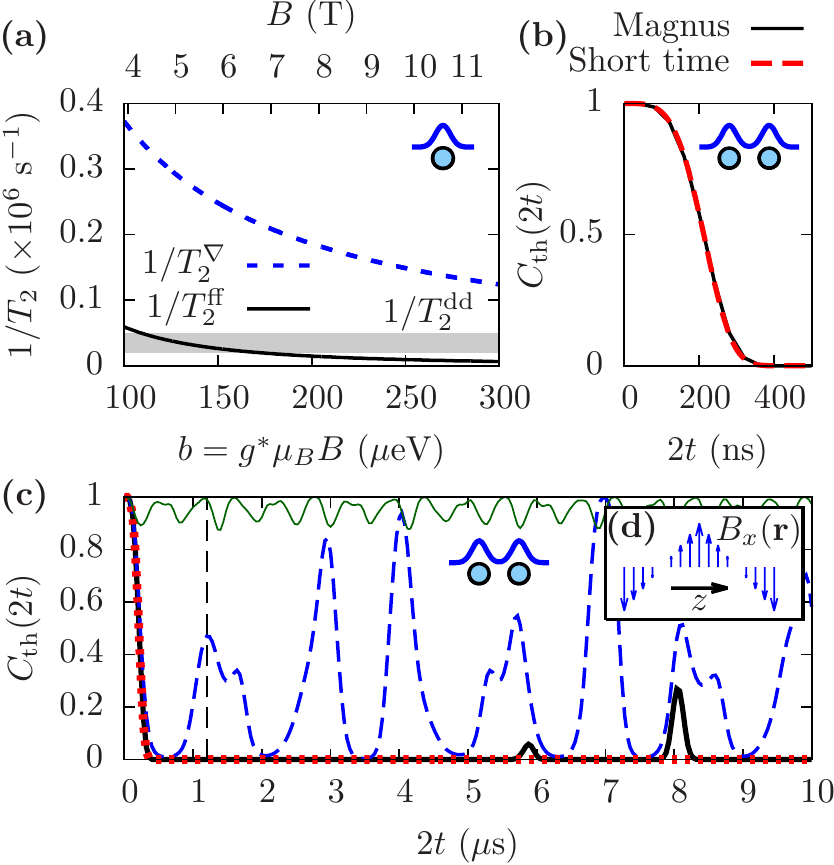}
	\end{center}
	\caption{(Color online) Dephasing due to a magnetic-field gradient in GaAs quantum dots. (a) Comparison of the hyperfine-induced dephasing times due to the flip-flop ($T_2^\mathrm{ff}$, black solid line) and nuclear dipolar interactions ($T_2^\mathrm{dd}$, gray region), with the dephasing time, $T_2^\nabla$, arising from the $\mvec B$-field gradient. These times have been evaluated for FID with a bath of $N=10^6$ nuclear spins that are prepared in a narrowed state. We consider a single electron spin in a single quantum dot with radius $r_0=40$ nm and $B_x(z)=z\left.\partial_zB^x\right|_{z=0}$, with $\left.\partial_zB^x\right|_{z=0}=1$ T$/\upmu$m, typical for experiments.\cite{pioro2008electrically,PhysRevB.81.085317} Solid black line: inverse dephasing time $1/T_2^\mathrm{ff}$ due to flip-flop hyperfine interaction.\cite{PhysRevB.77.125329} Shaded grey area: predicted inverse dephasing times $1/T_2^\mathrm{dd}$ due to nuclear dipole-dipole interactions, depending on dot size and shape.\cite{PhysRevB.74.035322} Dashed blue line: inverse dephasing time, $1/T_2^\nabla$, due to a magnetic-field gradient, from Eq.~\eq{eqn:T2nabla}. (b) Coherence factor for infinite-temperature thermal-state HE decay in a singlet-triplet qubit with dot radii $r_0=25$ nm and spacing $l=200$ nm. We also use $N=4.4\times10^6$, $B=45$~mT (parameters given in the experiment of Ref.~\onlinecite{bluhm2010dephasing}), and $\left.\partial_zB^x\right|_{z=0}=0.25$ T$/\upmu$m. The transverse magnetic field has the shape illustrated in Fig.~\ref{fig:1}. We compare the full leading-order Magnus expansion of Eq.~\eq{eqn:C:HE:th} (solid black line) to the short-time approximation of Eq.~\eq{eqn:sh} (dashed red line). (c) Full leading-order Magnus expansion results of Eq.~\eq{eqn:C:HE:th} for various $B$ values considered in the experiment of Ref.~\onlinecite{bluhm2010dephasing}: 45 mT (dotted red line), 95 mT (thick black line), 195 mT (dashed blue line), and 495 mT (thin green line). The dashed vertical line indicates the experimental dephasing time $T_{2,\mrm e}$, such that $C(2T_{2,e})=1/\mrm e$, measured for $B=45$~mT in Ref.~\onlinecite{bluhm2010dephasing}. Experimental dephasing times are longer for stronger magnetic fields. (d) Sawtooth-shaped transverse field distribution, for which leakage out of the $S$-$T_0$ subspace vanishes. \label{fig:GaAs}}
\end{figure}
We first check how the inverse decay time, $1/T_2^\nabla$, for an initially narrowed nuclear-spin bath under a magnetic-field gradient compares with the inverse dephasing times due to flip-flops ($1/T_2^\mathrm{ff}$) and nuclear-spin dipole-dipole couplings ($1/T_2^\mrm{dd}$). $1/T_2^\mathrm{ff}$ has been calculated from a Schrieffer-Wolff expansion\cite{PhysRevB.77.125329} for $b>A$, which for GaAs roughly corresponds to $B>4$ T. The derivations of Ref.~\onlinecite{PhysRevB.77.125329} result in $1/T_2^\mrm{ff}\propto 1/b^2$, while Eq.~\eq{eqn:T2nabla} predicts $1/T_2^\nabla\propto1/b$. Thus, for sufficiently large $b$, dephasing due to the gradient will always dominate over flip-flop decay. This is illustrated in Fig.~\ref{fig:GaAs}(a), where the dephasing rates are compared assuming typical parameters\cite{pioro2008electrically,PhysRevB.81.085317,PhysRevLett.107.146801} for a GaAs quantum dot and a magnetic-field gradient of $1\,\mathrm{T}/\mu\mathrm{m}$. In this particular case, $1/T_2^\nabla$ dominates over the entire magnetic-field range considered.

 In addition to flip-flops, nuclear dipolar interactions could also potentially lead to decay on a shorter time scale than the mechanisms considered here. Inverse dephasing times $1/T_2^\mrm{dd}$ due to dipole-dipole interactions have been predicted in Fig.~14 of Ref.~\onlinecite{PhysRevB.74.035322} to be between $10^4\,\mathrm{s}^{-1}$ and $10^5\,\mathrm{s}^{-1}$, with smaller values for small dot radii. For example, with a quantum-dot Bohr radius of $r_0=40$~nm, as in Fig.~\ref{fig:GaAs}(a), $1/T_2^\mrm{dd}$ lies between $2\times10^4$ s$^{-1}$ and $5\times10^4$ s$^{-1}$ [see Fig.~\ref{fig:GaAs}(a)] depending on the dot thickness and the crystal orientation. On the other hand, assuming $N\propto \pi r_0^2 y_0$, (where $y_0$ is the dot thickness), Eqs.~\eq{eqn:T2nabla} and~\eq{eqn:T2e} predict that $1/T_2^\nabla$ and $1/T_\mrm{2e}^\nabla$ are both independent of $r_0$.  Indeed, for single dots, $\Sigma^2\sim(A \delta b^x)^2/N\sim (A r_0\partial_z b^x)^2/N\sim1/y_0$, with $r_0$ contributions canceling out. Therefore, since $1/T_2^\mrm{ff}\propto 1/N\propto1/r_0^2$, the gradient dephasing mechanism dominates both flip-flop and dipole-dipole decay for sufficiently large dots with a fixed gradient, a situation that corresponds to the parameters of Fig.~\ref{fig:GaAs}(a), where $1/T_2^\nabla$ exceeds $1/T_2^\mathrm{ff},1/T_2^\mathrm{dd}$ by nearly an order of magnitude. Additionally, $1/T_2^\mrm{dd}$ is expected to be independent of $b$.\cite{yao2006theory} For large dots, gradient-induced dephasing should then dominate for a wide range of $b$. In summary, we have the following power-law hierarchy for the various mechanisms studied here: $1/T_2^\mrm{ff}\propto 1/b^2$, $1/T_2^\nabla\propto 1/b$, and $1/T_2^\nabla \propto 1$. For large dots it may then be possible to experimentally identify regimes where each mechanism dominates from a measurement of the inverse dephasing time as a function of $b$.

We now consider the case of GaAs singlet-triplet qubits, in which Hahn-echo dephasing dynamics without externally applied magnetic-field gradients has been extensively studied.\cite{petta2005coherent,bluhm2010dephasing} As explained in Appendix~\ref{sec:STMap}, the Hamiltonian for a system of two electron spins in the ${\{\ket S,\ket{T_0}\}}={\{(\ket{\upa\dwna}-\ket{\dwna\upa}/\sqrt2)},{(\ket{\upa\dwna}+\ket{\dwna\upa}/\sqrt2)\}}$ basis can be approximately mapped to the single-spin Hamiltonian of Eq.~\eq{eqn:H} when $b>\Delta b^x$. The results of Section~\ref{sec:thermal:HE} therefore apply equally well to the coherence of a singlet-triplet qubit [where, for a singlet-triplet qubit, $C(t)$ measures coherence between the states $\ket{\uparrow\downarrow}$ and $\ket{\downarrow\uparrow}$]. However, a finite value of $\Delta b^x\ne 0$ will typically lead to spin flips, resulting in leakage out of the ${\{\ket S,\ket{T_0}\}}$ subspace.  Here, $\Delta b^x=b^x_\mrm R-b^x_\mrm L$, where $b^x_\alpha\equiv \int d\mvec r \,g^*\mu_\mathrm{B} B^x(\mvec r)|\psi_\alpha(\mvec r)|^2$ and $\psi_{L(R)}(\mvec r)$ is a single-particle envelope function localized on the left (right) dot. The magnetic-field configuration illustrated in Fig.~\ref{fig:1}(a) leads, for example, to $\Delta b^x\ne 0$ and consequently to finite leakage due to spin flips.  One way to circumvent this leakage (as discussed below) would be to arrange a ``sawtooth" magnetic-field configuration, as shown in Fig. \ref{fig:GaAs}(d).

We now make predictions for Hahn-echo dephasing under a magnetic-field gradient, using parameters from an experiment by Bluhm \emph{et al.},\cite{bluhm2010dephasing} in which the Hahn-echo dephasing time for a singlet-triplet spin qubit in GaAs is measured as a function of a magnetic field $B$, with no applied gradient, $b_k^x=0$. We consider the effect of a weak transverse gradient $\left.\partial_zB^x\right|_{z=0}=0.25$ T$/\upmu$m, corresponding to $\Delta B^x=\Delta b^x/g^*\mu_\mathrm{B}=50$~mT. From Eq.~\eq{eqn:Bc}, with $\Sigma_s^2$ taken from Eq.~\eq{eqn:sigma2:double:ST0}, we estimate the critical field for motional averaging to be $B_c\sim225$~mT. Thus, it is possible to set $\Delta B^x<B<B_c$, such that leakage outside the $S$-$T_0$ subspace remains small, but loss of electron-spin coherence due to the gradient is observed. In that regime, as illustrated in Fig.~\ref{fig:GaAs}(b), the dynamics of the coherence factor is well-described by Eq.~\eq{eqn:sh}, which assumes a short-time expansion. Equation \eq{eqn:sh} predicts decay in a time $T_{2,e}^\nabla\simeq116~$ns, while the Hahn-echo measurements without the gradient yield a dephasing time $T_{2,\mrm e}\sim 600~\mathrm{ns}$ for $B=45$~mT. Thus, our model predicts that introducing a transverse magnetic-field gradient of only $0.25$~T$/\upmu$m in that experiment would severely decrease coherence times. However, as illustrated by Fig.~\ref{fig:GaAs}(c), taking $B\gg B_c$ would drive this system to a motional-averaging regime, avoiding rapid decay from the magnetic-field gradient. 

Finally, we stress that the results of the previous paragraph correspond to a best-case scenario for the geometry of Fig.~\ref{fig:1}(a) using a singlet-triplet qubit, where all leakage out of the computational subspace has been neglected. With the magnetic-field distribution $B^x(\mathbf{r})$ of Fig.~\ref{fig:1}(a), there would be no leakage due to spin flips for a bonding/antibonding molecular state $\psi_\pm(\mathbf{r})$ with equal weight on right and left dots, chosen such that $\int d^3 r |\psi_\pm(\mathbf{r})|^2B^x(\mathbf{r})=0$. For an electron in one of the states $\psi_\pm(\mathbf{r})$, $T_{2,e}^\nabla$ and $B_c$ would be comparable to the values obtained above assuming localized states in the left/right dot, $\psi_{L,R}(\mathbf{r})$.  Alternatively, localized orbital states $\psi_{L,R}(\mathbf{r})$ combined with the sawtooth field distribution $B^x(\mathbf{r})$ of Fig.~\ref{fig:GaAs}(d) would also avoid leakage provided $\int d^3 r |\psi_{L,R}(\mathbf{r})|^2B^x(\mathbf{r})=0$.

\subsection{Silicon}

We now apply our model to electron-spin qubits in Si, considering quantum dots and single P donors. Mainly because the density of nuclear spins is much lower than in GaAs, the hyperfine field in Si is much smaller. Indeed, in natural Si, which contains $4.7\%$ $^{29}$Si nuclei, we typically have\cite{PhysRevB.83.165301,maune2012coherent,schliemann2003electron} $A\sim100$~neV. Without a magnetic-field gradient, for the isotopic concentration of natural silicon, Witzel \textit{et al.} (see Fig. 23 of Ref.~\onlinecite{PhysRevB.86.035452}) have calculated a Hahn-echo dephasing time $T_{2,e}$ between 100 and 300~$\upmu$s in both quantum dots and single donors (including hyperfine interaction and dipole-dipole couplings)Ê. These values are also supported by experiments performed on phosphorus-doped silicon at low concentration.\cite{PhysRevB.68.193207,PhysRevB.82.121201} Neglecting dipole-dipole and flip-flop interactions, our model predicts dephasing times due to a moderate magnetic-field gradient that are shorter than this 100-$\upmu$s timescale. We will therefore neglect contributions to dephasing from flip-flop and dipolar interactions in the remainder of this section, focusing on the dominant magnetic-field-gradient mechanism.

\subsubsection{Silicon quantum dots\label{sec:lat:Si}}

\begin{figure}
	\begin{center}
		\includegraphics[scale=1]{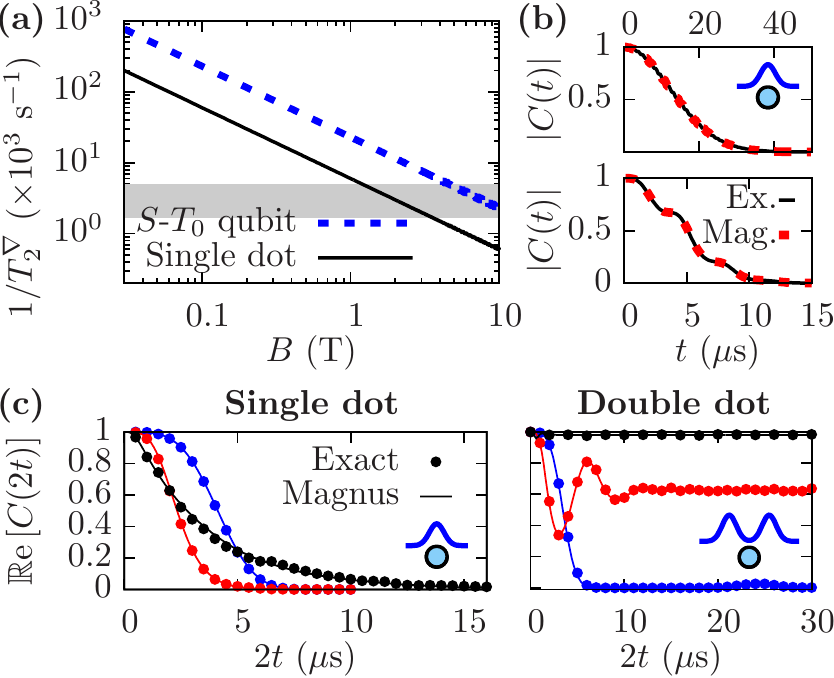}
	\end{center}
	\caption{(Color online) Dephasing due to a magnetic-field gradient in Si quantum dots. (a) Inverse decay times for narrowed-state FID from Eq.~\eq{eqn:T2nabla}. We take a magnetic-field gradient $\left.\partial_zB^x\right|_{z=0}=1$T$/\upmu$m and a quantum-dot Bohr radius $r_0=15$~nm, with inter-dot spacing $l=80$~nm in the double-dot case. We also take a total hyperfine coupling $A=210$~neV and $N=10^4$ from Refs.~\onlinecite{PhysRevB.83.165301,maune2012coherent}. Shaded gray area:  range of total HE inverse dephasing times $2T_{2,e}$ in Si without a magnetic field gradient.\cite{PhysRevB.86.035452} (b) Coherence dynamics for narrowed-state FID in a single dot. Top: $B=100$~mT, bottom: $B=30$~mT. Dashed red line: Magnus expansion results, from Eq.~\eq{eqn:C:int:FID:n}, solid black line: exact solution, from Eq.~\eq{eqn:C:FID}. (c) Coherence dynamics for infinite-temperature thermal-state HE, with $B\rightarrow0$. Solid lines: Magnus expansion prediction, from Eq.~\eq{eqn:C:HE:th}. Dots: exact solution, from Eq.~\eq{eqn:C:HE}. Blue: $\Delta B^x=20$~mT, red: $\Delta B^x=80$~mT, black: $\Delta B^x=400$~mT, with $\Delta B^x=\Delta b^x/g^*\mu_\mathrm{B}$ and $\Delta b^x$ defined in Section~\ref{sec:narrowed:FID}. \label{fig:Si}}
\end{figure}

We first discuss results for narrowed-state free-induction decay. Fig.~\ref{fig:Si}(a) shows $1/T_2^\nabla$ as predicted from Eq.~\eq{eqn:T2nabla} for $B$ ranging from $30$~mT to 10~T. For $B<100$~mT and for the parameters described in the caption of Fig.~\ref{fig:Si}, $T_2^\nabla$ is smaller than $10$~$\upmu$s, i.e.~at least an order of magnitude shorter than $2T_{2,e}$ taken from Ref.~\onlinecite{PhysRevB.86.035452}, illustrated by the shaded gray area. However, for $B>1$~T, $T_2^\nabla$ can be pushed beyond $100$~$\upmu$s, providing a way to avoid most of this additional dephasing. Fig.~\ref{fig:Si}(b) shows the coherence dynamics of a spin in a single quantum dot. Interestingly, for $b\sim\delta b^x$ (with $\delta b^x\equiv r_0\left.\partial_z b^x\right|_{z=0}$), the result deviates from pure Gaussian ($\sim\eul{-t^2}$) behavior. The coherence factor $C(t)$ displays additional envelope modulations at the nuclear-spin precession frequency, $\sim\gamma b$. This effect is illustrated in the lower plot of Fig.~\ref{fig:Si}(b). The agreement between the Magnus expansion and the exact solution is very good and even reproduces these small oscillations. Precise conditions for the validity of the Magnus expansion are derived in Appendix~\ref{sec:Magnus}. We find qualitatively similar results to those presented in Fig.~\ref{fig:Si}(b) for a single spin in a double dot.

Because of the smallness of $A$ in silicon, the critical gradients required for both motional averaging and exponential decay in HE decay that were obtained in Section~\ref{sec:thermal:HE} are much smaller than in GaAs. Indeed, for a single dot, exponential decay from the onset of a Markovian regime would require $\delta b^x/g^*\mu_\mathrm{B}>\delta b^x_\mathrm{M}/g^*\mu_\mathrm{B}\simeq2\,\mathrm{T}$ in GaAs, while in silicon such a regime is obtained for $\delta b^x/g^*\mu_\mathrm{B}>\delta b^x_\mathrm{M}/g^*\mu_\mathrm{B}\simeq20$~mT. Since flip-flop interactions are negligible in silicon dots even for $B\sim1$~mT, a transition from the non-Markovian to the Markovian regime could, in principle, be observed by tuning $\delta b^x/g^*\mu_B$ from $\sim20$~mT to a value $>100$~mT, as illustrated in the left plot of Fig.~\ref{fig:Si}(c), though sustaining such a large gradient on a length-scale of $\sim15\,$nm may pose a challenge with current technology. Additionally, the right-hand-side plot of Fig.~\ref{fig:Si}(c) shows that, consistent with Section~\ref{sec:thermal:HE}, the coherence properties of a double dot in the long-time limit are radically different from those of a single dot. Indeed, for large $\Delta B^x$, rather than reaching a Markovian regime, the gradient can induce motional averaging. Both the single- and double-dot calculations shown here are performed with $B=0$ to emphasize that the gradient itself is the source of that behavior, though calculations for a small but finite field of $B\sim1$~mT (required to justify the secular-hyperfine coupling assumption) give similar results.

\subsubsection{Phosphorus donors in silicon\label{sec:Si:P}}

We now turn to single phosphorus donor spin qubits in silicon~\cite{kane1998silicon,morello2010single,pla2012single} (Si:P). In this system, the logical qubit is the spin of a single electron bound to the phosphorus atom.  The P nucleus has a finite spin $I=1/2$. When $b$ is much larger than the strength of the coupling between the electron spin and the donor nuclear spin, $A_\mrm{SI}$, the system is well-described by the Hamiltonian of Eq.~\eq{eqn:H}. The relevant crossover magnetic field is $A_\mrm{SI}/(g^\ast\mu_B)\sim1$~mT, which can be achieved with a moderate applied magnetic field.\cite{kane1998silicon}

\begin{figure}
	\begin{center}
		\includegraphics[scale=1]{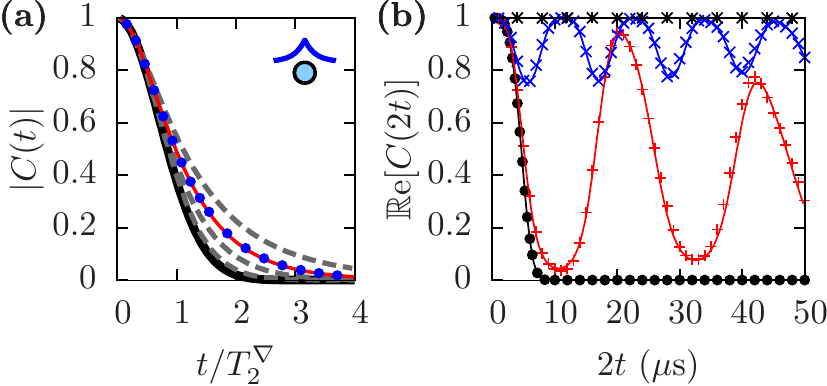}
	\end{center}
	\caption{(Color online) Coherence dynamics of an electron spin qubit in a single P donor in Si. We take $N=250$, $A=210$~neV, and $\left.\partial_{z}B_x\right|_{z=0}=1$~T/$\upmu$m. The system corresponds to a single dot with $r_0=3$~nm, $d=3$ and $q=1$. (a) Narrowed-state FID with $B=200$~mT. Thick black curve: prediction from the Magnus expansion under the Gaussian approximation, Eq.~\eq{eqn:C:int:FID:n}. Thin red curve: exact solution from Eq.~\eq{eqn:C:FID}. Blue dots: Magnus expansion without Gaussian approximation, from Eq.~\eq{eqn:nonGaussian}. Gray dashed lines, from top to bottom: exact coherence decay, respectively, for $N=125$, 500, and 2000. (b) Hahn-echo decay with initial thermal states for $N=250$, with varying $B$. From bottom to top: $B=0$, 10~mT, 20~mT and 100~mT. Points: exact solution of Eq.~\eq{eqn:C:HE}. Solid lines of corresponding color: prediction from the Magnus expansion, Eq.~\eq{eqn:C:HE:th}.
	\label{fig:SiP}}
\end{figure}

First, we calculate the evolution of coherence in narrowed-state FID. As shown in Fig.~\ref{fig:SiP}(a), Eq.~\eq{eqn:C:int:FID:n} (solid black line) predicts Gaussian decay with time constant $T_2^\nabla\simeq65$~$\upmu$s, much smaller than $2T_{2,e}\sim600\;\upmu$s
\footnote{
	In the absence of measured narrowed-state decay in Si, we compare the FID decay time $T_2^\nabla$ to the $1/\mrm e$-decay time 2$T_{2,e}$ for Hahn echo accounting for the full $2\tau$ duration between intialization and echo.
}
measured in the absence of a magnetic-field gradient.\cite{PhysRevB.68.193207,PhysRevB.82.121201}  However, the corresponding exact solution, calculated from Eq.~\eq{eqn:C:HE} and illustrated by the red line deviates from a Gaussian in the long-time limit. Indeed, after a short-lived Gaussian-dominated behavior, the exact solution exhibits an exponential tail. Yet, according to Eq.~\eq{eqn:criterion:Magnus:FID}, the Magnus expansion should converge rapidly for up to $\delta B^x = \delta b^x/g^*\mu_\mathrm{B}\sim600\,\mathrm{mT}$, which is overwhelmingly larger than $\delta B^x=3$~mT, as chosen here. We thus turn to the other approximation introduced in the derivation of Eq.~\eq{eqn:C:int:FID:n}: the Gaussian approximation.

To rigorously check that non-Gaussian contributions are responsible for this divergence in behavior, we notice that for $I=\frac12$, which is the case for a bath of $^{29}$Si nuclei, the exponential in Eq.~\eq{eqn:exp:exact}, $e^{i\delta_\chi \hat{X}(\tau)}$, can be expanded using the identity
\be
	\eul{i\mvec v\cdot\hat{\mvec\sigma}}=\cos v+i\frac{\mvec v\cdot\hat{\mvec\sigma}}{v}\sin v,
\ee
where $\mvec v$ is an arbitrary vector and $\hat{\mvec\sigma}$ is the vector of Pauli matrices $(\sx,\sy,\sz)$. Doing so, the coherence factor can be expressed as a product,
\be
	C_j(t)=\eul{ibt+i\sum_kh_k^zm_k^j}\prod_{k=1}^\infty\left(\cos\frac{|\mvec h_k|}{4}+4i\frac{h_k^zm_k^j}{|\mvec h_k|}\sin\frac{|\mvec h_k|}{4}\right).\label{eqn:nonGaussian}
\ee
We thus find an expression for the coherence factor under the leading-order Magnus expansion that includes non-Gaussian corrections if we replace $\mvec h_k(t)$ by its value given in Eq.~\eq{eqn:h:FID}. The result for single donors is given by the blue dots in Fig.~\ref{fig:SiP}(a). Strikingly, the leading-order Magnus expansion overlaps with the exact solution, clearly showing the breakdown of the Gaussian approximation for small nuclear-spin baths.

Though finding a criterion for the validity of the Gaussian approximation directly from Eq.~\eq{eqn:nonGaussian} is not a straightforward task, an order-of-magnitude estimate is sufficient here.  As explained in Appendix~\ref{sec:Gaussian}, non-Gaussian corrections are suppressed only as a weak power of $N$ ($\sim 1/N^{1/4}$).  Thus, $N$ must be fairly large for the Gaussian approximation to work, a few hundreds typically not being sufficient. Indeed, for $N=250$, we have $N^{1/4}\simeq3.97\sim$1. As illustrated by the dashed gray lines in Fig.~\ref{fig:SiP}, as $N$ increases, the exponential tail due to the non-Gaussian corrections fades away, with dephasing becoming almost entirely Gaussian for $N\sim2000$. To generate this plot, as $N$ is increased, we have adjusted $A$ to keep the ratio $N^{5/6}\gamma b/A$ constant, such that the criterion for convergence of the Magnus expansion expressed by Eq.~\eq{eqn:criterion:Magnus:FID} remains satisfied.

Finally, Fig.~\ref{fig:SiP}(b) gives the evolution of the coherence factor under Hahn echo for a nuclear-spin bath initially in an infinite-temperature thermal state. From Eq.~\eq{eqn:Bc}, we find that the critical magnetic field for the onset of motional averaging is $B_c\sim19$~mT. This is confirmed by the exact solution of Eq~\eq{eqn:C:HE}, as shown by the points in Fig.~\ref{fig:SiP}(b). In past experiments,\cite{pla2012single} an external field $B\sim1$~T has been used. For magnetic-field gradients of $\sim1$~T/$\upmu$m, these experiments would be deep in the motional averaging regime, making dephasing due to the magnetic-field gradient completely reversible by Hahn echo. On a separate note, for Hahn echo, the predictions from the combination of Magnus expansion and Gaussian approximation closely fit with the exact solution, as illustrated by the solid lines. This does not contradict the results of Appendix~\ref{sec:Gaussian}, since the criteria found there are sufficient but not necessary conditions for non-Gaussian contributions to be negligible. 

Finally, from Eq.~\eq{eqn:C:HE:th}, we obtain a criterion for Markovian Hahn-echo decay due to a magnetic-field gradient that is very similar to Eq.~\eq{eqn:deltabx:markov}, except that $A_k$ values are taken for $d=3$, $q=1$. We find that Markovian behavior in single donors requires $\delta B^x\sim100$~mT. With $r_0=3$~nm, this would require a gradient $\left.\partial_zB^x\right|_{z=0}>30$~T/$\upmu$m, which could be technically very difficult to achieve. Larger quantum dots such as those discussed in Section~\ref{sec:lat:Si} would thus be more appropriate to observe that behavior.

\section{Conclusions}\label{sec:Conclusions}

We have calculated the free-induction decay and Hahn-echo dynamics of an electron spin interacting with a nuclear-spin bath in the presence of an inhomogeneous magnetic field. Accounting for only secular hyperfine coupling, we have given exact solutions for spin coherence in terms of a product of $\mathcal{O}(N)$ matrices.  Additionally, we have found closed-form simple analytical expressions within a leading-order Magnus expansion and Gaussian approximation.

In the case of free-induction decay, the coherence factor typically decays as $\sim\eul{-t^2}$. In the case of Hahn echo, we have found $\sim\eul{-t^4}$ dephasing if the longitudinal magnetic field is below a known threshold, $B\lesssim B_c$, above which this system enters a motional-averaging regime characterized by an incomplete decay of spin coherence. 
For electron spins in single quantum dots, a large transverse magnetic-field gradient decreases the nuclear-spin correlation time, leading to Markovian (exponential) rather than $\sim \eul{-t^4}$ echo dephasing. In contrast, for spins in double quantum dots we rather obtain motional averaging and an associated incomplete coherence decay in the limit of a large uniform magnetic-field gradient.

We have further investigated the relevance of the above results to three physical systems: quantum dots in GaAs or Si and single P donors in Si. In each case, the inverse dephasing time due to the magnetic-field gradient mechanism $1/T_2^\nabla$ can be much larger than those due to electron-nuclear flip-flop ($1/T_2^\mathrm{ff}$) and nuclear dipole-dipole interactions ($1/T_2^\mathrm{dd}$), i.e.~$1/T_2^\nabla\gg 1/T_2^\mathrm{ff}, 1/T_2^\mathrm{dd}$. This dominance of $1/T_2^\nabla$ can be confirmed in GaAs singlet-triplet qubits [see Fig.~\ref{fig:GaAs}(c)]. Additionally, we have found that single quantum dots in Si would be best suited to observe a cross-over between non-Markovian and Markovian dephasing. Finally, we have shown that the Gaussian approximation fails to correctly predict decoherence in small systems such as single P donors in Si. This result highlights the difference between spin baths and bosonic baths, for which Wick's theorem allows for exact suppression of all non-Gaussian contributions.

The model developed here applies directly to heavy-hole spin qubits, for which flip-flop terms can be suppressed through confinement and the secular-hyperfine limit can be reached even for very weak external magnetic fields.\cite{PhysRevLett.109.237601} Our results may also give insight into dephasing in spin-orbit qubits in InAs nanowires,\cite{nadj2010spin} where strain leads to inhomogeneous nuclear quadrupolar interactions that may decorrelate the nuclear-spin system as in the case of a magnetic-field gradient.

\begin{acknowledgments}
 We thank M.~Pioro-Ladri\`ere and X.~Wang for useful discussions. We acknowledge financial support from NSERC, CIFAR, FRQNT, INTRIQ, and the W.~C.~Sumner Foundation.
\end{acknowledgments}

\appendix

\section{Hyperfine coupling constants in a double quantum dot\label{sec:A}}

In this Appendix, we obtain expressions for the hyperfine coupling strengths $A_k$ for three possible geometries: (i) one electron in a single dot; (ii) one electron in the (bonding or antibonding) delocalized molecular state of a symmetric double quantum dot; (iii) two electrons in a double quantum dot (each in the localized $\ket L$ and $\ket R$ states), in the $S$-$T_0$ basis. Fig.~\ref{fig:1} helps to visualize the problem in two dimensions.

Case (i) has already been treated in the literature,\cite{PhysRevB.77.125329} yielding for an electron wavefunction of the form $\psi(r)=\psi(0)\exp[-(r/r_0)^q/2]$, in dimension $d$,
\be
	A_k=Av_0|\psi(\mvec r_k)|^2=\frac{A}{N\frac dq\Gamma\left(\frac dq\right)}\eul{-(k/N)^{q/d}},\label{eqn:Ak:single}
\ee
with $v_0$ the volume per nuclear spin and $N$ the number of nuclei within the dot radius $r_0$.

We deal with the two cases involving a double dot in a very similar way, using left ($\sigma=L$) and right ($\sigma=R$) basis states $\psi_\sigma(\vecr_k)=\psi(\vecr^\sigma)\exp[-(r_k^\sigma/r_0)^q/2]$, where $\vecr_k^\sigma=\vecr_k-\vecr^\sigma$ and the vectors $\vecr^\sigma$ locate the center of each dot. Using spherical coordinates, $(r_k^\sigma)^2=r_k^2+l^2/4 +r_k l\,G_{d,k}^\sigma$, with
\be
	G_{d,k}^{L/R}=\left\{\begin{array}{ll} 
			\pm(-1)^k&d=1,\\
			\pm\cos\theta_k&d=2,\\
			\pm\sin\theta_k\cos\varphi_k&d=3.
		\end{array}\right.\label{eqn:G:k:d}
\ee
For $d=2$, $\theta_k$ is the polar angle that locates the nuclear spin $k$, while for $d=3$, $\theta_k$ and $\varphi_k$ are, respectively, the polar and azimuthal angles. As in Ref.~\onlinecite{PhysRevB.77.125329}, we define $N$ as the number of nuclear spins within the Bohr radius $r_0$ and label the spins in increasing order of $r_k$, such that $r_k/r_0=(k/N)^{1/d}$. Thus, for an electron in dot $\sigma$, the hyperfine couplings are given by
\be
	A_k^{\sigma}=A_0^{\sigma}\exp\left\{-\left[\left(\frac{k}{N}\right)^{2/d}+\eta^2+2\eta\left(\frac{k}{N}\right)^{1/d}G_{d,k}^{\sigma}\right]^{q/2}\right\},
\ee
with $\eta\equiv l/2r_0$, the dimensionless separation of the double dot, and $A_0^\sigma$ is obtained from the normalization condition $\sum_kA_k^\sigma=A$.

In the case of a single electron shared by two dots of the same size and at the same electrical potential, the electron wavefunction is $\psi_\pm(\vecr)=\frac{1}{\sqrt2}[\psi_L(\vecr)\pm\psi_R(\vecr)]$, treating both symmetric ($+$) and antisymmetric ($-$) superpositions at the same time. In order to calculate $A_0^\sigma$ and thus obtain $A_k^\pm$, we assume an isotropic distribution of nuclear spins and $N\gg 1$ and evaluate $A=\sum_k A_k^\pm $ by converting $\sum_k$ to an integral to obtain for $d=q=2$
\be
	A_k^\pm=Av_0|\psi_\pm(\mvec r_k)|^2\simeq\frac{A}{N}\frac{\cosh\left(2\eta\sqrt{\frac{k}{N}}\cos\theta_k\right)\pm1}{\eul{\eta^2}\pm1}\eul{-k/N}.\label{eqn:Ak:double:1e}
\ee
In the above, we have defined $A^\pm_k$ as the hyperfine coupling strength for nuclear spin $k$ when the electron wavefunction is $\psi_\pm(\mvec r)$. As expected, $A_k^+$ tends toward the single-dot expression [Eq. \eqref{eqn:Ak:single}] for $\eta\propto l\rightarrow0$.

Hyperfine couplings for case (iii) are obtained in a similar way, however it is first required to map the Hamiltonian of a singlet-triplet qubit to Eq.~\eq{eqn:H}. This is done in Appendix~\ref{sec:STMap}.

\section{Mapping to a singlet-triplet qubit\label{sec:STMap}}

For moderate electron Zeeman spittings $b>\Delta b^x$, the model of Eq.~\eq{eqn:H} can be applied to describe two electron spins in the $S$-$T_0$ basis,\cite{PhysRevB.72.125337} defined by $\ket{S/T_0}=\frac{1}{\sqrt2}({\ket{\upa\dwna}}\mp{\ket{\dwna\upa}})$. Indeed, defining $\hat{\tau}_x=\ket{S}\bra{T_0}+\ket{T_0}\bra S$ and $\hat{\tau}_z=\proj{T_0}-\proj S$ and assuming a negligible exchange coupling, the effective Hamiltonian describing the two electron spins and their nuclear-spin bath in a magnetic-field gradient is
\be
	\hat H_\mrm{ST_0}=\hat{H}_\mrm Z^z+\hat{H}_\mrm Z^x+\delta \hat{h}^z\hat{\tau}^x+b\sum_k\gamma_k\hat{I}_k^z+\sum_k\gamma_kb_k^x\Ikx,\label{eqn:HST0}
\ee
where $\hat H_\mrm Z^\alpha$ is the Zeeman coupling of the electrons for a magnetic field along direction $\alpha$. Defining $\ket{T_+}=\ket{\upa\upa}$ and $\ket{T_-}=\ket{\dwna\dwna}$, we find
\begin{align}
	\hat H_\mrm Z^z&=b\left(\proj{T_+}-\proj{T_-}\right),\\
	\hat H_\mrm Z^x&=b^x\frac{\ket{T_+}+\ket{T_-}}{\sqrt2}\bra{T_0}-\frac{\Delta b^x}{2}\frac{\ket{T_+}-\ket{T_-}}{\sqrt2}\bra S+\mrm{H.c.},\notag
\end{align}
where $b^x\equiv(b_\mrm R^x+b_\mrm L^x)/2$, $\Delta b^x\equiv b_\mrm R^x-b_\mrm L^x$ and $b_i^x$ is the electron Zeeman splitting along $x$ in the localized orbital state of dot $i$. For a magnetic field $B^x(z)$ that is an odd function of $z$, as shown in Fig.~\ref{fig:1}, $b^x=0$. The term $\propto \Delta b^x$ above typically does not vanish, and thus may lead to leakage out of the $S$-$T_0$ subspace. This leakage can, however, be suppressed if $b\gg\Delta b^x$. In this limit, we neglect the term $\propto \Delta b^x$ and take $\hat\tau^x\rightarrow2\Sz$ in the Hamiltonian of Eq.~\eq{eqn:HST0}, which approximately maps it to Eq.~\eq{eqn:H}, provided we replace $A_k\rightarrow\delta A_k=A_k^L-A_k^R$. The coherence factor then corresponds to the off-diagonal element of the density matrix in the $\left\{\ket{\upa\dwna},\ket{\dwna\upa}\right\}$ basis. Using the results of Appendix~\ref{sec:A}, we find ($d=q=2$)
\be
	\delta A_k=2\frac{A}{N}\eul{-\left(\frac{k}{N}+\eta^2\right)}\sinh\left(2\eta\sqrt{\frac{k}{N}}\cos\theta_k\right).\label{eqn:Ak:double:ST0}
\ee

We estimate corrections to the above mapping for finite $\Delta b^x$ from the leakage probability, $1-\mean{\hat P_{\mrm{ST_0}}(t)}$, where $\hat P_{\mrm{ST_0}}\equiv\proj{S}+\proj{T_0}$.  We have evaluated $\mean{\hat P_{\mrm{ST_0}}(t)}$ within time-dependent perturbation theory to leading order in $\Delta b^x/b$ and $A_k/\omega_k$. This calculation predicts the onset of a stationary leakage probability $\sim (\Delta b^x/b)^2$ after a time $\sim\sqrt{N}/A$. In principle, this probability could grow over time if other processes that can decorrelate the electron and nuclear spins were included, such as nuclear dipolar interactions. In that case, we expect the leakage probability to grow on a time scale $\sim (b/\Delta b^x)^2\tau_c$, with $\tau_c$ the correlation time of the nuclear-spin bath. In bulk GaAs, $\tau_c\sim100~\mu$s can be estimated from the NMR linewidth.\cite{Paget1977low} We therefore expect the approximate mapping to give an accurate representation of dynamics in the singlet-triplet basis for times $t \lesssim 100\,\mu\mathrm{s}$ whenever $\Delta b^x/b\lesssim 1$.

\section{Validity of the Magnus expansion\label{sec:Magnus}}

Using the Gaussian approximation, we showed in Section~\ref{sec:gauss} that the coherence factor is of the form $C(\tau)\sim\exp[\sum_k|\mvec h_k(\tau)|^2]$, as expressed in Eq.~\eq{eqn:C:th} and Eq.~\eq{eqn:C:n}. The field $\mvec h_k(t)$ is then obtained from the Hamiltonian of Eq.~\eq{eqn:H} with a zeroth-order Magnus expansion. To subleading order in that expansion, we would have, for $\alpha\in\{x,y,z\}$
\begin{align}
	\left(h_k^\alpha\right)^2&\sim\left[h_k^{\alpha(0)}\right]^2+2h_k^{\alpha(0)}h_k^{\alpha(2)}+\left[h_k^{\alpha(1)}\right]^2.
\end{align}
In the rest of this Appendix, we drop the $\alpha$ index to ease notation. In the above expression, $h_k^{(n)}\sim(A_kb_k^x/\omega_k)^{n+1}$, such that $h_k^{(0)}\sim(A_kb_k^x/\omega_k)^{2}$ is the leading term, considered in the main body of the paper, while $h_k^{(0)}h_k^{(2)}$ and $[h_k^{(1)}]^2$, of order $(A_kb_k^x/\omega_k)^{4}$, are subleading. Note that we have dropped terms containing odd powers of $A_kb_k^x$ because they cancel out when summed over $k$ due to the symmetries of $|\psi(\mvec r)|^2$ and $B^x(\mvec r)$. We will now establish criteria for the leading term to dominate the subleading ones, and hence for the Magnus expansion to be valid.

The $n$-th order term in the Magnus expansion involves $n+1$ integrals over time of $\mvec v_k(t)$, which is given by Eq.~\eq{eqn:vk}. Roughly speaking, each integral of a sine generates a factor $\cos\omega_kt/\omega_k$ in $h_k^{(n)}$, while each integral of a cosine generates a factor $t\sin\omega_kt$. Thus, we find
\begin{align}
	h_k^{(0)}&\sim\frac{\gamma A_kb_k^x}{\omega_k^2}+\frac{\gamma A_kb_k^xt}{\omega_k},\;\;\; h_k^{(1)}\sim\left(\frac{\gamma A_kb_k^x}{\omega_k}\right)^2\frac{t}{\omega_k},\\
	h_k^{(2)}&\sim\frac{A_k^3}{\omega_k^2}\left(\frac{\gamma b_k^x}{\omega_k}\right)^3t+\frac{A_k^3}{\omega_k}\left(\frac{\gamma b_k^x}{\omega_k}\right)^3t^2,
\end{align}
such that 
\begin{align}
	\left[h_k^{(1)}\right]^2+h_k^{(0)}h_k^{(2)}&\sim\frac{A_k^4}{\omega_k^3}\left(\frac{\gamma b_k^x}{\omega_k}\right)^4t+\frac{A_k^4}{\omega_k^2}\left(\frac{\gamma b_k^x}{\omega_k}\right)^4t^2\notag\\
	&\qquad+\frac{A_k^4}{\omega_k}\left(\frac{\gamma b_k^x}{\omega_k}\right)^4t^3.
\end{align}
Using an order-of-magnitude estimate of the sum over $k$, this gives three critical times $(t_1, t_2, t_3)$ for which each of the above terms can be of order 1. In the limit where $b\gg\Delta b_x$, we find
\begin{align}
	t_1\sim\frac{\gamma^3N^3b^7}{(A\Delta b^x)^4},\; t_2\sim\frac{\gamma N^{3/2}b^3}{(A\Delta b^x)^2},\;t_3\sim\frac{\gamma^{1/3}Nb^{5/3}}{(A\Delta b^x)^{4/3}},\label{eqn:timescale:large:B}
\end{align}
while in the opposite limit ($b\ll\Delta b^x$), we get
\begin{align}
	t_1\sim\frac{(\gamma N\Delta b^x)^3}{A^4},\;t_2\sim\frac{\gamma N^{3/2}\Delta b^x}{A^2},\;t_3\sim\frac{N(\gamma \Delta b^x)^{1/3}}{A^{4/3}}.\label{eqn:timescale:small:B}
\end{align}
If $t_\mrm{crit}=\mrm{min}\{t_i\}$ is shorter than the predicted dephasing time from the leading-order Magnus expansion, we conclude that the approximation is not valid. This approach is especially convenient to fix a validity condition for narrowed-state free-induction decay, since there is no motional averaging in that regime. For $b\gg\Delta b^x$, comparing $T_2^\nabla$ from Eq.~\eq{eqn:T2nabla} to the above three timescales sets three criteria on $\Delta b^x$, of which the most stringent is
\be
	\lambda\equiv \frac{\Delta b^x}{b}<\frac{N^{5/6}\gamma b}{A}.\label{eqn:criterion:Magnus:FID}
\ee
Considering a worst-case scenario with $\lambda\sim1$, this criterion becomes $b>A/\gamma N^{5/6}=b_\mrm{min}^\mrm M$. Such minimum values $B_\mathrm{min}^M = b_\mrm{min}^\mrm M/g^*\mu_\mathrm{B}$ in the three main classes of devices studied here are displayed in Table~\ref{tab:FID}. These numbers allow us to conclude that keeping only the leading order in the Magnus expansion is a good approximation in all the free-induction cases considered here.
\begin{table}
	\begin{center}
		\begin{ruledtabular}
			\begin{tabular}{ccccc}
				Material	& $A$ (MHz)	& $\gamma$ (MHz/T)	& $N$	& $B_\mrm{min}^\mrm M$\\
				\hline
				GaAs	& $1.3\times10^5$	& 60	& $4.4\times10^6$	& $50$ mT\\
				Si		& $320$			& 53	& $10^4$			& $0.3$ mT\\
				Si:P		& $320$			& 53	& $250$			& $60$ mT\\
			\end{tabular}
		\end{ruledtabular}
	\end{center}
	\caption{Minimum longitudinal fields $B_\mrm{min}^\mrm{M}$ below which the Magnus expansion can fail to describe narrowed-state FID.
\label{tab:FID}}
\end{table}

\begin{table}
	\begin{center}
		\begin{ruledtabular}
			\begin{tabular}{lccc}
				\multicolumn{4}{l}{\textbf{Small longitudinal field limit} ($b\ll\Delta b^x$)}\\
				GaAs	& $\Delta B^x=25$ mT	& $\Delta B^x=100$ mT	& $\Delta B^x=200$ mT\\
				$t_\mrm{crit}$	& $800$ ns	& $1.2$ $\upmu$s	& $1.5$ $\upmu$s\\
				Si		& $\Delta B^x=20$ mT	& $\Delta B^x=80$ mT	& $\Delta B^x=400$ mT\\
				$t_\mrm{crit}$	& $4.5$ $\upmu$s	& $7.5$ $\upmu$s	& $13$ $\upmu$s\\	
				\hline
				\multicolumn{4}{l}{\textbf{Large longitudinal field limit} ($b\gg\Delta b^x$)}\\	
				GaAs	& $B=200$ mT	& $B=295$ mT	& $B=695$ mT\\
				$t_\mrm{crit}$	& $1.5$ $\upmu$s	& 3 $\upmu$s	& 12 $\upmu$s\\
				Si:P		& $B=10$~mT	& $B=20$~mT	& $B=100$~mT\\
				$t_\mrm{crit}$	& $0.5$ $\upmu$s 	& $1.5$~$\upmu$s	& 20~$\upmu$s
			\end{tabular}
		\end{ruledtabular}
	\end{center}
	\caption{Timescales $t_\mrm{crit}$ below which the Magnus expansion is valid, according to Eqs.~\eq{eqn:timescale:large:B} and~\eq{eqn:timescale:small:B}. Unless specified otherwise in the table, parameters for lateral dots are those taken from Figs.~\ref{fig:GaAs} and~\ref{fig:Si}, considering double dots.
\label{tab:HE}}
\end{table}
For Hahn-echo decay, it is less convenient to fix a boundary on parameters, because motional averaging can keep coherence finite for long times. Thus, we directly study the  timescales beyond which the Magnus expansion can fail. These timescales are calculated in dots and donor impurities for the various parameter regimes studied in this paper and are presented in Table~\ref{tab:HE}. In situations where decay occurs in a finite time $T_\mrm{2e}^\nabla$, that decay time is always shorter than $t_\mrm{crit}$, except in single donors. In situations where motional averaging is predicted, no discrepancy between the Magnus expansion and the exact numerical solution is seen. We emphasize that the analysis done here only gives us critical times $t_\mrm{crit}$ below which the leading-order term in the Magnus expansion is sure to dominate, but does not determine where the subleading-order term becomes dominant. In other words, we believe these estimates give a definite range of applicability, but the approximations introduced may well be valid outside of that range.

\section{Validity of the Gaussian approximation\label{sec:Gaussian}}

In this Appendix, we find a simple criterion which, if respected, justifies the Gaussian approximation. In Section~\ref{sec:gauss}, we found that the coherence factor is related to $\mean{\eul{i\delta \hat X(\tau)}}$, which can be expanded in its Taylor series
\be
	\mean{\eul{i\delta \hat X}}=\sum_n\frac{i^n}{n!}\sum_{k_1...k_n}\sum_{l_1...l_n}h_{k_1}^{l_1}...h_{k_n}^{l_n}\mean{\delta \hat I_{k_1}^{l_1}...\delta \hat I_{k_n}^{l_n}}.
\ee
Since we have taken a bath for which nuclei are uncorrelated, we have $\mean{\delta \hat I_{k_1}^{l_1}\delta \hat I_{k_2}^{l_2}}=\mean{\delta \hat I_{k_1}^{l_1}}\mean{\delta \hat I_{k_2}^{l_2}}=0\;\forall\;k_1\neq k_2$, by definition of $\delta \hat I_k$. To simplify the notation, in this Appendix, we define $\delta\hat X\equiv\delta_\chi\hat X$ and $\delta\hat{\mvec{I}}_k\equiv\delta_\chi\hat{\mathbf{I}}_k$ and drop the $\chi$ index in expectation values. Thus, we have
\begin{align}
	\mean{\eul{i\delta \hat X(\tau)}}&\simeq\sum_n\frac{i^{2n}}{(2n)!}a_n\sum_{kll'}\prod_{q=1}^n h_{k_q}^{l_q}h_{k_q}^{l_q'}\mean{\delta \hat I_{k_q}^{l_q}\delta \hat I_{k_q}^{l_q'}},
\end{align}
with $a_n=(2n)!/2^nn!$ the number of ways to group $2n$ terms into $n$ pairs. In the above, we have dropped all moments of order greater than two, which lead to non-Gaussian contributions. The above sums can then be reorganized to yield
\begin{align}
	\mean{\eul{i\delta \hat X(\tau)}}&\simeq\exp\left[-\frac12\sum_{kll'}h_k^l(\tau)h_k^{l'}(\tau)\mean{\delta \hat I_k^l\delta \hat I_k^{l'}}\right],
\end{align}
which is the Gaussian approximation used in Section~\ref{sec:gauss} and throughout the paper.

To justify the omission of the moments of order larger than two, we invoke the large number $N$ of nuclei. Indeed, there are roughly $N$ times less elements in a sum of the form $\sum_{k_1...k_n}$ over moments of order 3 than in a similar sum over moments of order 2. Furthermore, in both thermal and narrowed states, we find that moments of odd order vanish. Thus, in order to find the conditions under which the Gaussian approximation is justified, we estimate the importance of the sum over fourth-order moments. Considering only products of these fourth-order moments, we find that the non-Gaussian contributions are of the order of
\be
	C_{4}(\tau)\sim\sum_n\frac{i^{4n}b_n}{(4n)!}\sum\prod_{q=1}^n h_{k_q}^{l_q}h_{k_q}^{l_q'}h_{k_q}^{l_q''}h_{k_q}^{l_q'''}\mean{\delta \hat I_{q_1}^{q_1}\delta \hat I_{k_q}^{l_q'}\delta \hat I_{k_q}^{l_q''}\delta \hat I_{k_q}^{l_q'''}},
\ee
with $b_n=(4n)!/24^nn!$ the number of ways to group $4n$ terms into strings of four. For a nuclear-spin bath initially in a narrowed state, this leads us to
\be
	C_4(t)\sim\exp\left\{\frac{1}{24}\sum_k\left[(h_k^x(t))^4+(h_k^y(t))^4\right]\right\}.
\ee
For $b\gg\Delta b^x$, this becomes
\be
	C_4(t)\sim\exp\left[\sum_k\left(\frac{A_kb_k^xt}{b}\right)^4\right].
\ee
Thus, non-Gaussian contributions grow like $\exp[(t/\tau_4)^4]$, with $1/\tau_4\sim A\Delta b^x/N^{3/4}b$, while Gaussian contributions led to $\exp[-(t/T_2^\nabla)^2]$ decay. Comparing $1/\tau_4$ and $1/T_2^\nabla\sim A\Delta b^x/N^{1/2}b$, we find that $\tau_4/T_2^\nabla\sim N^{1/4}$, and thus that non-Gaussian contributions only decay as a weak power law as $N$ is increased.

In the case of Hahn echo, we rather find 
\be
	C_4(2t)\sim\exp\left\{\sum_{k\alpha}[h_k^\alpha(t)]^4\right\}.
\ee
Taking $b\ll\Delta b^x$, if the short-time expansion is valid, we can expand the sines and cosines in the fields $h_k^\alpha(t)$ to leading order in $t$. We obtain $C_4(2t)\sim\exp[(t/\tau_8)^8]$, and we find $\tau_8/T_\mrm{2e}^\nabla\sim N^{1/8}$. Another relevant limit is when $b\gg\Delta b^x$, useful to describe motional averaging. We can then replace $\omega_k\rightarrow b$ and the argument of the exponential becomes an oscillating function with amplitude $\sim (A_kb_k^x)^4/b^8$. Requiring this term to be smaller than the equivalent amplitude for the Gaussian term, discussed in Section~\ref{sec:thermal:HE}, we find that the following criterion must be met
\be
	\lambda=\frac{\Delta b^x}{b}<\frac{\sqrt N \gamma b}{A}.
\ee
This criterion is easily satisfied in all materials studied here, even in single P donor impurities in Si, in accordance with the fact that the Gaussian approximation does not fail to predict motional averaging in that material [see Fig.~\ref{fig:SiP}(b)].

\section{$\Sigma_s^2$ for relevant geometries \label{sec:Sigma2}}

As shown in Section~\ref{sec:spec}, both the free-induction and Hahn-echo dephasing times can be linked to the sum $\Sigma_s^2=\sum_\ks(A_\ks b_\ks^x)^2$. In this section, we calculate $\Sigma_s^2$ for various geometries. To simplify the notation, we will drop the species index $s$.

In all cases, we assume a transverse magnetic field of the form $B^x(\vecr)=\beta z$, with $\beta$ a constant, such that
\begin{align}
	b_k^x=z_k\left.\partial_zb^x\right|_{z=0}=\frac{r_k\cos\theta_k}{r_0}\delta b^x=\sqrt{\frac kN}\delta b^x\cos\theta_k,\label{eqn:bkx}
\end{align}
where $\theta_k$ is the polar angle locating nucleus $k$ and nuclei are labeled with increasing $r_k$, as in Appendix~\ref{sec:A}. We have also defined $\delta b^x\equiv r_0\left.\partial_zb^x\right|_{z=0}$.

Using Eqs.~\eq{eqn:Ak:single} and~\eq{eqn:bkx}, we calculate $\Sigma^2$ for a single dot assuming an isotropic distribution of nuclear spins, with $N$ large enough to convert the sum into an integral. We obtain
\be
	\Sigma^2=\frac{1}{d}\frac{(A\delta b^x)^2}{N\frac dq2^{\frac{2+d}{q}}}\frac{\Gamma\left(\frac{2+d}{q}\right)}{\Gamma^2\left(\frac{d}{q}\right)}.\label{eqn:sigma2:single}
\ee
In the specific case of $d=q=2$, this yields $\Sigma^2=(A\delta b^x)^2/8N$.

For double dots with $d=q=2$, we rather use Eqs.~\eq{eqn:Ak:double:1e} and~\eq{eqn:Ak:double:ST0} for $A_k$. In the case of a single electron with a symmetric ($+$) or antisymmetric ($-$) delocalized wavefunction $\psi_\pm(\mvec r)$ as in Appendix~\ref{sec:A}, we get
\be
	\Sigma^2=\left(\frac{A\delta b^x}{\eul{\eta^2}\pm1}\right)^2\frac{3\pm4\eul{\eta^2/2}(1+\eta^2)+\eul{2\eta^2}(1+4\eta^2)}{16N}.\label{eqn:sigma2:double:1e}
\ee
Different wavefunctions lead to different values of $\Sigma^2$ because we have taken into account the overlap between the orbitals. For a singlet-triplet qubit, we rather have
\be
	\Sigma^2=(A\delta b^x)^2\frac{1+4\eta^2-\eul{-2\eta^2}}{4N}.\label{eqn:sigma2:double:ST0}
\ee


\bibliography{articleReply}

\begin{thebibliography}{50}
\expandafter\ifx\csname natexlab\endcsname\relax\def\natexlab#1{#1}\fi
\expandafter\ifx\csname bibnamefont\endcsname\relax
  \def\bibnamefont#1{#1}\fi
\expandafter\ifx\csname bibfnamefont\endcsname\relax
  \def\bibfnamefont#1{#1}\fi
\expandafter\ifx\csname citenamefont\endcsname\relax
  \def\citenamefont#1{#1}\fi
\expandafter\ifx\csname url\endcsname\relax
  \def\url#1{\texttt{#1}}\fi
\expandafter\ifx\csname urlprefix\endcsname\relax\def\urlprefix{URL }\fi
\providecommand{\bibinfo}[2]{#2}
\providecommand{\eprint}[2][]{\url{#2}}

\bibitem[{\citenamefont{Koppens et~al.}(2008)\citenamefont{Koppens, Nowack, and
  Vandersypen}}]{PhysRevLett.100.236802}
\bibinfo{author}{\bibfnamefont{F.~H.~L.} \bibnamefont{Koppens}},
  \bibinfo{author}{\bibfnamefont{K.~C.} \bibnamefont{Nowack}},
  \bibnamefont{and} \bibinfo{author}{\bibfnamefont{L.~M.~K.}
  \bibnamefont{Vandersypen}}, \bibinfo{journal}{Phys. Rev. Lett.}
  \textbf{\bibinfo{volume}{100}}, \bibinfo{pages}{236802}
  (\bibinfo{year}{2008}),
  \urlprefix\url{http://link.aps.org/doi/10.1103/PhysRevLett.100.236802}.

\bibitem[{\citenamefont{Bluhm et~al.}(2010{\natexlab{a}})\citenamefont{Bluhm,
  Foletti, Neder, Rudner, Mahalu, Umansky, and Yacoby}}]{bluhm2010dephasing}
\bibinfo{author}{\bibfnamefont{H.}~\bibnamefont{Bluhm}},
  \bibinfo{author}{\bibfnamefont{S.}~\bibnamefont{Foletti}},
  \bibinfo{author}{\bibfnamefont{I.}~\bibnamefont{Neder}},
  \bibinfo{author}{\bibfnamefont{M.}~\bibnamefont{Rudner}},
  \bibinfo{author}{\bibfnamefont{D.}~\bibnamefont{Mahalu}},
  \bibinfo{author}{\bibfnamefont{V.}~\bibnamefont{Umansky}}, \bibnamefont{and}
  \bibinfo{author}{\bibfnamefont{A.}~\bibnamefont{Yacoby}},
  \bibinfo{journal}{Nature Physics} \textbf{\bibinfo{volume}{7}},
  \bibinfo{pages}{109} (\bibinfo{year}{2010}{\natexlab{a}}).

\bibitem[{\citenamefont{Khaetskii et~al.}(2002)\citenamefont{Khaetskii, Loss,
  and Glazman}}]{PhysRevLett.88.186802}
\bibinfo{author}{\bibfnamefont{A.~V.} \bibnamefont{Khaetskii}},
  \bibinfo{author}{\bibfnamefont{D.}~\bibnamefont{Loss}}, \bibnamefont{and}
  \bibinfo{author}{\bibfnamefont{L.}~\bibnamefont{Glazman}},
  \bibinfo{journal}{Phys. Rev. Lett.} \textbf{\bibinfo{volume}{88}},
  \bibinfo{pages}{186802} (\bibinfo{year}{2002}),
  \urlprefix\url{http://link.aps.org/doi/10.1103/PhysRevLett.88.186802}.

\bibitem[{\citenamefont{Maune et~al.}(2012)\citenamefont{Maune, Borselli,
  Huang, Ladd, Deelman, Holabird, Kiselev, Alvarado-Rodriguez, Ross, Schmitz
  et~al.}}]{maune2012coherent}
\bibinfo{author}{\bibfnamefont{B.}~\bibnamefont{Maune}},
  \bibinfo{author}{\bibfnamefont{M.}~\bibnamefont{Borselli}},
  \bibinfo{author}{\bibfnamefont{B.}~\bibnamefont{Huang}},
  \bibinfo{author}{\bibfnamefont{T.}~\bibnamefont{Ladd}},
  \bibinfo{author}{\bibfnamefont{P.}~\bibnamefont{Deelman}},
  \bibinfo{author}{\bibfnamefont{K.}~\bibnamefont{Holabird}},
  \bibinfo{author}{\bibfnamefont{A.}~\bibnamefont{Kiselev}},
  \bibinfo{author}{\bibfnamefont{I.}~\bibnamefont{Alvarado-Rodriguez}},
  \bibinfo{author}{\bibfnamefont{R.}~\bibnamefont{Ross}},
  \bibinfo{author}{\bibfnamefont{A.}~\bibnamefont{Schmitz}},
  \bibnamefont{et~al.}, \bibinfo{journal}{Nature}
  \textbf{\bibinfo{volume}{481}}, \bibinfo{pages}{344} (\bibinfo{year}{2012}).

\bibitem[{\citenamefont{Coish et~al.}(2008)\citenamefont{Coish, Fischer, and
  Loss}}]{PhysRevB.77.125329}
\bibinfo{author}{\bibfnamefont{W.~A.} \bibnamefont{Coish}},
  \bibinfo{author}{\bibfnamefont{J.}~\bibnamefont{Fischer}}, \bibnamefont{and}
  \bibinfo{author}{\bibfnamefont{D.}~\bibnamefont{Loss}},
  \bibinfo{journal}{Phys. Rev. B} \textbf{\bibinfo{volume}{77}},
  \bibinfo{pages}{125329} (\bibinfo{year}{2008}),
  \urlprefix\url{http://link.aps.org/doi/10.1103/PhysRevB.77.125329}.

\bibitem[{\citenamefont{Coish et~al.}(2010)\citenamefont{Coish, Fischer, and
  Loss}}]{PhysRevB.81.165315}
\bibinfo{author}{\bibfnamefont{W.~A.} \bibnamefont{Coish}},
  \bibinfo{author}{\bibfnamefont{J.}~\bibnamefont{Fischer}}, \bibnamefont{and}
  \bibinfo{author}{\bibfnamefont{D.}~\bibnamefont{Loss}},
  \bibinfo{journal}{Phys. Rev. B} \textbf{\bibinfo{volume}{81}},
  \bibinfo{pages}{165315} (\bibinfo{year}{2010}),
  \urlprefix\url{http://link.aps.org/doi/10.1103/PhysRevB.81.165315}.

\bibitem[{\citenamefont{Witzel and Das~Sarma}(2006)}]{PhysRevB.74.035322}
\bibinfo{author}{\bibfnamefont{W.~M.} \bibnamefont{Witzel}} \bibnamefont{and}
  \bibinfo{author}{\bibfnamefont{S.}~\bibnamefont{Das~Sarma}},
  \bibinfo{journal}{Phys. Rev. B} \textbf{\bibinfo{volume}{74}},
  \bibinfo{pages}{035322} (\bibinfo{year}{2006}),
  \urlprefix\url{http://link.aps.org/doi/10.1103/PhysRevB.74.035322}.

\bibitem[{\citenamefont{Cywi\ifmmode~\acute{n}\else \'{n}\fi{}ski
  et~al.}(2009{\natexlab{a}})\citenamefont{Cywi\ifmmode~\acute{n}\else
  \'{n}\fi{}ski, Witzel, and Das~Sarma}}]{PhysRevLett.102.057601}
\bibinfo{author}{\bibfnamefont{L.}~\bibnamefont{Cywi\ifmmode~\acute{n}\else
  \'{n}\fi{}ski}}, \bibinfo{author}{\bibfnamefont{W.~M.} \bibnamefont{Witzel}},
  \bibnamefont{and}
  \bibinfo{author}{\bibfnamefont{S.}~\bibnamefont{Das~Sarma}},
  \bibinfo{journal}{Phys. Rev. Lett.} \textbf{\bibinfo{volume}{102}},
  \bibinfo{pages}{057601} (\bibinfo{year}{2009}{\natexlab{a}}),
  \urlprefix\url{http://link.aps.org/doi/10.1103/PhysRevLett.102.057601}.

\bibitem[{\citenamefont{Cywi\ifmmode~\acute{n}\else \'{n}\fi{}ski
  et~al.}(2009{\natexlab{b}})\citenamefont{Cywi\ifmmode~\acute{n}\else
  \'{n}\fi{}ski, Witzel, and Das~Sarma}}]{PhysRevB.79.245314}
\bibinfo{author}{\bibfnamefont{L.}~\bibnamefont{Cywi\ifmmode~\acute{n}\else
  \'{n}\fi{}ski}}, \bibinfo{author}{\bibfnamefont{W.~M.} \bibnamefont{Witzel}},
  \bibnamefont{and}
  \bibinfo{author}{\bibfnamefont{S.}~\bibnamefont{Das~Sarma}},
  \bibinfo{journal}{Phys. Rev. B} \textbf{\bibinfo{volume}{79}},
  \bibinfo{pages}{245314} (\bibinfo{year}{2009}{\natexlab{b}}),
  \urlprefix\url{http://link.aps.org/doi/10.1103/PhysRevB.79.245314}.

\bibitem[{\citenamefont{Cywi\ifmmode~\acute{n}\else \'{n}\fi{}ski
  et~al.}(2010)\citenamefont{Cywi\ifmmode~\acute{n}\else \'{n}\fi{}ski,
  Dobrovitski, and Das~Sarma}}]{PhysRevB.82.035315}
\bibinfo{author}{\bibfnamefont{L.}~\bibnamefont{Cywi\ifmmode~\acute{n}\else
  \'{n}\fi{}ski}}, \bibinfo{author}{\bibfnamefont{V.~V.}
  \bibnamefont{Dobrovitski}}, \bibnamefont{and}
  \bibinfo{author}{\bibfnamefont{S.}~\bibnamefont{Das~Sarma}},
  \bibinfo{journal}{Phys. Rev. B} \textbf{\bibinfo{volume}{82}},
  \bibinfo{pages}{035315} (\bibinfo{year}{2010}),
  \urlprefix\url{http://link.aps.org/doi/10.1103/PhysRevB.82.035315}.

\bibitem[{\citenamefont{Barnes et~al.}(2012)\citenamefont{Barnes,
  Cywi\ifmmode~\acute{n}\else \'{n}\fi{}ski, and Das~Sarma}}]{Barnes2012}
\bibinfo{author}{\bibfnamefont{E.}~\bibnamefont{Barnes}},
  \bibinfo{author}{\bibfnamefont{L.}~\bibnamefont{Cywi\ifmmode~\acute{n}\else
  \'{n}\fi{}ski}}, \bibnamefont{and}
  \bibinfo{author}{\bibfnamefont{S.}~\bibnamefont{Das~Sarma}},
  \bibinfo{journal}{Phys. Rev. Lett.} \textbf{\bibinfo{volume}{109}},
  \bibinfo{pages}{140403} (\bibinfo{year}{2012}),
  \urlprefix\url{http://link.aps.org/doi/10.1103/PhysRevLett.109.140403}.

\bibitem[{\citenamefont{Klauder and Anderson}(1962)}]{Klauder1962}
\bibinfo{author}{\bibfnamefont{J.~R.} \bibnamefont{Klauder}} \bibnamefont{and}
  \bibinfo{author}{\bibfnamefont{P.~W.} \bibnamefont{Anderson}},
  \bibinfo{journal}{Phys. Rev.} \textbf{\bibinfo{volume}{125}},
  \bibinfo{pages}{912} (\bibinfo{year}{1962}),
  \urlprefix\url{http://link.aps.org/doi/10.1103/PhysRev.125.912}.

\bibitem[{\citenamefont{Witzel et~al.}(2005)\citenamefont{Witzel, de~Sousa, and
  Das~Sarma}}]{Witzel2005}
\bibinfo{author}{\bibfnamefont{W.~M.} \bibnamefont{Witzel}},
  \bibinfo{author}{\bibfnamefont{R.}~\bibnamefont{de~Sousa}}, \bibnamefont{and}
  \bibinfo{author}{\bibfnamefont{S.}~\bibnamefont{Das~Sarma}},
  \bibinfo{journal}{\prb} \textbf{\bibinfo{volume}{72}},
  \bibinfo{pages}{161306} (\bibinfo{year}{2005}),
  \urlprefix\url{http://link.aps.org/doi/10.1103/PhysRevB.72.161306}.

\bibitem[{\citenamefont{Yao et~al.}(2006)\citenamefont{Yao, Liu, and
  Sham}}]{yao2006theory}
\bibinfo{author}{\bibfnamefont{W.}~\bibnamefont{Yao}},
  \bibinfo{author}{\bibfnamefont{R.~B.} \bibnamefont{Liu}}, \bibnamefont{and}
  \bibinfo{author}{\bibfnamefont{L.~J.} \bibnamefont{Sham}},
  \bibinfo{journal}{\prb} \textbf{\bibinfo{volume}{74}},
  \bibinfo{pages}{195301} (\bibinfo{year}{2006}).

\bibitem[{\citenamefont{Morello et~al.}(2010)\citenamefont{Morello, Pla,
  Zwanenburg, Chan, Tan, Huebl, M{\"o}tt{\"o}nen, Nugroho, Yang, van Donkelaar
  et~al.}}]{morello2010single}
\bibinfo{author}{\bibfnamefont{A.}~\bibnamefont{Morello}},
  \bibinfo{author}{\bibfnamefont{J.}~\bibnamefont{Pla}},
  \bibinfo{author}{\bibfnamefont{F.}~\bibnamefont{Zwanenburg}},
  \bibinfo{author}{\bibfnamefont{K.}~\bibnamefont{Chan}},
  \bibinfo{author}{\bibfnamefont{K.}~\bibnamefont{Tan}},
  \bibinfo{author}{\bibfnamefont{H.}~\bibnamefont{Huebl}},
  \bibinfo{author}{\bibfnamefont{M.}~\bibnamefont{M{\"o}tt{\"o}nen}},
  \bibinfo{author}{\bibfnamefont{C.}~\bibnamefont{Nugroho}},
  \bibinfo{author}{\bibfnamefont{C.}~\bibnamefont{Yang}},
  \bibinfo{author}{\bibfnamefont{J.}~\bibnamefont{van Donkelaar}},
  \bibnamefont{et~al.}, \bibinfo{journal}{Nature}
  \textbf{\bibinfo{volume}{467}}, \bibinfo{pages}{687} (\bibinfo{year}{2010}).

\bibitem[{\citenamefont{Pla et~al.}(2012)\citenamefont{Pla, Tan, Dehollain,
  Lim, Morton, Jamieson, Dzurak, and Morello}}]{pla2012single}
\bibinfo{author}{\bibfnamefont{J.}~\bibnamefont{Pla}},
  \bibinfo{author}{\bibfnamefont{K.}~\bibnamefont{Tan}},
  \bibinfo{author}{\bibfnamefont{J.}~\bibnamefont{Dehollain}},
  \bibinfo{author}{\bibfnamefont{W.}~\bibnamefont{Lim}},
  \bibinfo{author}{\bibfnamefont{J.}~\bibnamefont{Morton}},
  \bibinfo{author}{\bibfnamefont{D.}~\bibnamefont{Jamieson}},
  \bibinfo{author}{\bibfnamefont{A.}~\bibnamefont{Dzurak}}, \bibnamefont{and}
  \bibinfo{author}{\bibfnamefont{A.}~\bibnamefont{Morello}},
  \bibinfo{journal}{Nature} \textbf{\bibinfo{volume}{489}},
  \bibinfo{pages}{541} (\bibinfo{year}{2012}).

\bibitem[{\citenamefont{Pioro-Ladri\`ere
  et~al.}(2008)\citenamefont{Pioro-Ladri\`ere, Obata, Tokura, Shin, Kubo,
  Yoshida, Taniyama, and Tarucha}}]{pioro2008electrically}
\bibinfo{author}{\bibfnamefont{M.}~\bibnamefont{Pioro-Ladri\`ere}},
  \bibinfo{author}{\bibfnamefont{T.}~\bibnamefont{Obata}},
  \bibinfo{author}{\bibfnamefont{Y.}~\bibnamefont{Tokura}},
  \bibinfo{author}{\bibfnamefont{Y.}~\bibnamefont{Shin}},
  \bibinfo{author}{\bibfnamefont{T.}~\bibnamefont{Kubo}},
  \bibinfo{author}{\bibfnamefont{K.}~\bibnamefont{Yoshida}},
  \bibinfo{author}{\bibfnamefont{T.}~\bibnamefont{Taniyama}}, \bibnamefont{and}
  \bibinfo{author}{\bibfnamefont{S.}~\bibnamefont{Tarucha}},
  \bibinfo{journal}{Nature Physics} \textbf{\bibinfo{volume}{4}},
  \bibinfo{pages}{776} (\bibinfo{year}{2008}).

\bibitem[{\citenamefont{Obata et~al.}(2010)\citenamefont{Obata,
  Pioro-Ladri\`ere, Tokura, Shin, Kubo, Yoshida, Taniyama, and
  Tarucha}}]{PhysRevB.81.085317}
\bibinfo{author}{\bibfnamefont{T.}~\bibnamefont{Obata}},
  \bibinfo{author}{\bibfnamefont{M.}~\bibnamefont{Pioro-Ladri\`ere}},
  \bibinfo{author}{\bibfnamefont{Y.}~\bibnamefont{Tokura}},
  \bibinfo{author}{\bibfnamefont{Y.-S.} \bibnamefont{Shin}},
  \bibinfo{author}{\bibfnamefont{T.}~\bibnamefont{Kubo}},
  \bibinfo{author}{\bibfnamefont{K.}~\bibnamefont{Yoshida}},
  \bibinfo{author}{\bibfnamefont{T.}~\bibnamefont{Taniyama}}, \bibnamefont{and}
  \bibinfo{author}{\bibfnamefont{S.}~\bibnamefont{Tarucha}},
  \bibinfo{journal}{Phys. Rev. B} \textbf{\bibinfo{volume}{81}},
  \bibinfo{pages}{085317} (\bibinfo{year}{2010}),
  \urlprefix\url{http://link.aps.org/doi/10.1103/PhysRevB.81.085317}.

\bibitem[{\citenamefont{Hu et~al.}(2012)\citenamefont{Hu, Liu, and
  Nori}}]{PhysRevB.86.035314}
\bibinfo{author}{\bibfnamefont{X.}~\bibnamefont{Hu}},
  \bibinfo{author}{\bibfnamefont{Y.-x.} \bibnamefont{Liu}}, \bibnamefont{and}
  \bibinfo{author}{\bibfnamefont{F.}~\bibnamefont{Nori}},
  \bibinfo{journal}{Phys. Rev. B} \textbf{\bibinfo{volume}{86}},
  \bibinfo{pages}{035314} (\bibinfo{year}{2012}),
  \urlprefix\url{http://link.aps.org/doi/10.1103/PhysRevB.86.035314}.

\bibitem[{\citenamefont{Frey et~al.}(2012)\citenamefont{Frey, Leek, Beck,
  Blais, Ihn, Ensslin, and Wallraff}}]{PhysRevLett.108.046807}
\bibinfo{author}{\bibfnamefont{T.}~\bibnamefont{Frey}},
  \bibinfo{author}{\bibfnamefont{P.~J.} \bibnamefont{Leek}},
  \bibinfo{author}{\bibfnamefont{M.}~\bibnamefont{Beck}},
  \bibinfo{author}{\bibfnamefont{A.}~\bibnamefont{Blais}},
  \bibinfo{author}{\bibfnamefont{T.}~\bibnamefont{Ihn}},
  \bibinfo{author}{\bibfnamefont{K.}~\bibnamefont{Ensslin}}, \bibnamefont{and}
  \bibinfo{author}{\bibfnamefont{A.}~\bibnamefont{Wallraff}},
  \bibinfo{journal}{Phys. Rev. Lett.} \textbf{\bibinfo{volume}{108}},
  \bibinfo{pages}{046807} (\bibinfo{year}{2012}),
  \urlprefix\url{http://link.aps.org/doi/10.1103/PhysRevLett.108.046807}.

\bibitem[{\citenamefont{Peterssson et~al.}(2012)\citenamefont{Peterssson,
  McFaul, Schroer, Jung, Taylor, Houck, and Petta}}]{petersson2012circuit}
\bibinfo{author}{\bibfnamefont{K.~D.} \bibnamefont{Peterssson}},
  \bibinfo{author}{\bibfnamefont{L.~W.} \bibnamefont{McFaul}},
  \bibinfo{author}{\bibfnamefont{M.~D.} \bibnamefont{Schroer}},
  \bibinfo{author}{\bibfnamefont{M.}~\bibnamefont{Jung}},
  \bibinfo{author}{\bibfnamefont{J.~M.} \bibnamefont{Taylor}},
  \bibinfo{author}{\bibfnamefont{A.~A.} \bibnamefont{Houck}}, \bibnamefont{and}
  \bibinfo{author}{\bibfnamefont{J.~R.} \bibnamefont{Petta}},
  \bibinfo{journal}{Nature} \textbf{\bibinfo{volume}{490}},
  \bibinfo{pages}{380} (\bibinfo{year}{2012}).

\bibitem[{\citenamefont{Trif et~al.}(2008)\citenamefont{Trif, Golovach, and
  Loss}}]{PhysRevB.77.045434}
\bibinfo{author}{\bibfnamefont{M.}~\bibnamefont{Trif}},
  \bibinfo{author}{\bibfnamefont{V.~N.} \bibnamefont{Golovach}},
  \bibnamefont{and} \bibinfo{author}{\bibfnamefont{D.}~\bibnamefont{Loss}},
  \bibinfo{journal}{Phys. Rev. B} \textbf{\bibinfo{volume}{77}},
  \bibinfo{pages}{045434} (\bibinfo{year}{2008}),
  \urlprefix\url{http://link.aps.org/doi/10.1103/PhysRevB.77.045434}.

\bibitem[{\citenamefont{Coish and Loss}(2004)}]{Coish2004}
\bibinfo{author}{\bibfnamefont{W.~A.} \bibnamefont{Coish}} \bibnamefont{and}
  \bibinfo{author}{\bibfnamefont{D.}~\bibnamefont{Loss}},
  \bibinfo{journal}{Phys. Rev. B} \textbf{\bibinfo{volume}{70}},
  \bibinfo{pages}{195340} (\bibinfo{year}{2004}),
  \urlprefix\url{http://link.aps.org/doi/10.1103/PhysRevB.70.195340}.

\bibitem[{\citenamefont{Klauser et~al.}(2006)\citenamefont{Klauser, Coish, and
  Loss}}]{PhysRevB.73.205302}
\bibinfo{author}{\bibfnamefont{D.}~\bibnamefont{Klauser}},
  \bibinfo{author}{\bibfnamefont{W.~A.} \bibnamefont{Coish}}, \bibnamefont{and}
  \bibinfo{author}{\bibfnamefont{D.}~\bibnamefont{Loss}},
  \bibinfo{journal}{Phys. Rev. B} \textbf{\bibinfo{volume}{73}},
  \bibinfo{pages}{205302} (\bibinfo{year}{2006}),
  \urlprefix\url{http://link.aps.org/doi/10.1103/PhysRevB.73.205302}.

\bibitem[{\citenamefont{Stepanenko et~al.}(2006)\citenamefont{Stepanenko,
  Burkard, Giedke, and Imamoglu}}]{Stepanenko2006enhancement}
\bibinfo{author}{\bibfnamefont{D.}~\bibnamefont{Stepanenko}},
  \bibinfo{author}{\bibfnamefont{G.}~\bibnamefont{Burkard}},
  \bibinfo{author}{\bibfnamefont{G.}~\bibnamefont{Giedke}}, \bibnamefont{and}
  \bibinfo{author}{\bibfnamefont{A.}~\bibnamefont{Imamoglu}},
  \bibinfo{journal}{\prl} \textbf{\bibinfo{volume}{96}},
  \bibinfo{pages}{136401} (\bibinfo{year}{2006}).

\bibitem[{\citenamefont{Giedke et~al.}(2006)\citenamefont{Giedke, Taylor,
  D'Alessandro, Lukin, and Imamo{\u{g}}lu}}]{Giedke2006quantum}
\bibinfo{author}{\bibfnamefont{G.}~\bibnamefont{Giedke}},
  \bibinfo{author}{\bibfnamefont{J.~M.} \bibnamefont{Taylor}},
  \bibinfo{author}{\bibfnamefont{D.}~\bibnamefont{D'Alessandro}},
  \bibinfo{author}{\bibfnamefont{M.~D.} \bibnamefont{Lukin}}, \bibnamefont{and}
  \bibinfo{author}{\bibfnamefont{A.}~\bibnamefont{Imamo{\u{g}}lu}},
  \bibinfo{journal}{\pra} \textbf{\bibinfo{volume}{74}},
  \bibinfo{pages}{032316} (\bibinfo{year}{2006}).

\bibitem[{\citenamefont{Bluhm et~al.}(2010{\natexlab{b}})\citenamefont{Bluhm,
  Foletti, Mahalu, Umansky, and Yacoby}}]{PhysRevLett.105.216803}
\bibinfo{author}{\bibfnamefont{H.}~\bibnamefont{Bluhm}},
  \bibinfo{author}{\bibfnamefont{S.}~\bibnamefont{Foletti}},
  \bibinfo{author}{\bibfnamefont{D.}~\bibnamefont{Mahalu}},
  \bibinfo{author}{\bibfnamefont{V.}~\bibnamefont{Umansky}}, \bibnamefont{and}
  \bibinfo{author}{\bibfnamefont{A.}~\bibnamefont{Yacoby}},
  \bibinfo{journal}{Phys. Rev. Lett.} \textbf{\bibinfo{volume}{105}},
  \bibinfo{pages}{216803} (\bibinfo{year}{2010}{\natexlab{b}}),
  \urlprefix\url{http://link.aps.org/doi/10.1103/PhysRevLett.105.216803}.

\bibitem[{\citenamefont{Fischer et~al.}(2008)\citenamefont{Fischer, Coish,
  Bulaev, and Loss}}]{fischer2008spin}
\bibinfo{author}{\bibfnamefont{J.}~\bibnamefont{Fischer}},
  \bibinfo{author}{\bibfnamefont{W.~A.} \bibnamefont{Coish}},
  \bibinfo{author}{\bibfnamefont{D.~V.} \bibnamefont{Bulaev}},
  \bibnamefont{and} \bibinfo{author}{\bibfnamefont{D.}~\bibnamefont{Loss}},
  \bibinfo{journal}{Phys. Rev. B} \textbf{\bibinfo{volume}{78}},
  \bibinfo{pages}{155329} (\bibinfo{year}{2008}),
  \urlprefix\url{http://link.aps.org/doi/10.1103/PhysRevB.78.155329}.

\bibitem[{\citenamefont{Eble et~al.}(2009)\citenamefont{Eble, Testelin,
  Desfonds, Bernardot, Balocchi, Amand, Miard, Lema{\^ i}tre, Marie, and
  Chamarro}}]{Eble2009hole}
\bibinfo{author}{\bibfnamefont{B.}~\bibnamefont{Eble}},
  \bibinfo{author}{\bibfnamefont{C.}~\bibnamefont{Testelin}},
  \bibinfo{author}{\bibfnamefont{P.}~\bibnamefont{Desfonds}},
  \bibinfo{author}{\bibfnamefont{F.}~\bibnamefont{Bernardot}},
  \bibinfo{author}{\bibfnamefont{A.}~\bibnamefont{Balocchi}},
  \bibinfo{author}{\bibfnamefont{T.}~\bibnamefont{Amand}},
  \bibinfo{author}{\bibfnamefont{A.}~\bibnamefont{Miard}},
  \bibinfo{author}{\bibfnamefont{A.}~\bibnamefont{Lema{\^ i}tre}},
  \bibinfo{author}{\bibfnamefont{X.}~\bibnamefont{Marie}}, \bibnamefont{and}
  \bibinfo{author}{\bibfnamefont{M.}~\bibnamefont{Chamarro}},
  \bibinfo{journal}{\prl} \textbf{\bibinfo{volume}{102}},
  \bibinfo{pages}{146601} (\bibinfo{year}{2009}).

\bibitem[{\citenamefont{Brunner et~al.}(2009)\citenamefont{Brunner, Gerardot,
  Dalgarno, W{\"u}st, Karrai, Stoltz, Petroff, and
  Warburton}}]{Brunner2009coherent}
\bibinfo{author}{\bibfnamefont{D.}~\bibnamefont{Brunner}},
  \bibinfo{author}{\bibfnamefont{B.~D.} \bibnamefont{Gerardot}},
  \bibinfo{author}{\bibfnamefont{P.~A.} \bibnamefont{Dalgarno}},
  \bibinfo{author}{\bibfnamefont{G.}~\bibnamefont{W{\"u}st}},
  \bibinfo{author}{\bibfnamefont{K.}~\bibnamefont{Karrai}},
  \bibinfo{author}{\bibfnamefont{N.~G.} \bibnamefont{Stoltz}},
  \bibinfo{author}{\bibfnamefont{P.~M.} \bibnamefont{Petroff}},
  \bibnamefont{and} \bibinfo{author}{\bibfnamefont{R.~J.}
  \bibnamefont{Warburton}}, \bibinfo{journal}{Science}
  \textbf{\bibinfo{volume}{325}}, \bibinfo{pages}{70} (\bibinfo{year}{2009}).

\bibitem[{\citenamefont{De~Greve et~al.}(2011)\citenamefont{De~Greve, McMahon,
  Press, Ladd, Bisping, Schneider, Kamp, Worschech, H{\"o}fling, Forchel
  et~al.}}]{DeGreve2011ultrafast}
\bibinfo{author}{\bibfnamefont{K.}~\bibnamefont{De~Greve}},
  \bibinfo{author}{\bibfnamefont{P.~L.} \bibnamefont{McMahon}},
  \bibinfo{author}{\bibfnamefont{D.}~\bibnamefont{Press}},
  \bibinfo{author}{\bibfnamefont{T.~D.} \bibnamefont{Ladd}},
  \bibinfo{author}{\bibfnamefont{D.}~\bibnamefont{Bisping}},
  \bibinfo{author}{\bibfnamefont{C.}~\bibnamefont{Schneider}},
  \bibinfo{author}{\bibfnamefont{M.}~\bibnamefont{Kamp}},
  \bibinfo{author}{\bibfnamefont{L.}~\bibnamefont{Worschech}},
  \bibinfo{author}{\bibfnamefont{S.}~\bibnamefont{H{\"o}fling}},
  \bibinfo{author}{\bibfnamefont{A.}~\bibnamefont{Forchel}},
  \bibnamefont{et~al.}, \bibinfo{journal}{Nature Physics}
  \textbf{\bibinfo{volume}{7}}, \bibinfo{pages}{872} (\bibinfo{year}{2011}).

\bibitem[{\citenamefont{Wang et~al.}(2012)\citenamefont{Wang, Chesi, and
  Coish}}]{PhysRevLett.109.237601}
\bibinfo{author}{\bibfnamefont{X.~J.} \bibnamefont{Wang}},
  \bibinfo{author}{\bibfnamefont{S.}~\bibnamefont{Chesi}}, \bibnamefont{and}
  \bibinfo{author}{\bibfnamefont{W.~A.} \bibnamefont{Coish}},
  \bibinfo{journal}{Phys. Rev. Lett.} \textbf{\bibinfo{volume}{109}},
  \bibinfo{pages}{237601} (\bibinfo{year}{2012}),
  \urlprefix\url{http://link.aps.org/doi/10.1103/PhysRevLett.109.237601}.

\bibitem[{\citenamefont{Nadj-Perge et~al.}(2010)\citenamefont{Nadj-Perge,
  Frolov, Bakkers, and Kouwenhoven}}]{nadj2010spin}
\bibinfo{author}{\bibfnamefont{S.}~\bibnamefont{Nadj-Perge}},
  \bibinfo{author}{\bibfnamefont{S.}~\bibnamefont{Frolov}},
  \bibinfo{author}{\bibfnamefont{E.}~\bibnamefont{Bakkers}}, \bibnamefont{and}
  \bibinfo{author}{\bibfnamefont{L.}~\bibnamefont{Kouwenhoven}},
  \bibinfo{journal}{Nature} \textbf{\bibinfo{volume}{468}},
  \bibinfo{pages}{1084} (\bibinfo{year}{2010}).

\bibitem[{\citenamefont{Fermi}(1930)}]{Fermi:1930fk}
\bibinfo{author}{\bibfnamefont{E.}~\bibnamefont{Fermi}},
  \bibinfo{journal}{Zeitschrift f{\"u}r Physik} \textbf{\bibinfo{volume}{60}},
  \bibinfo{pages}{320} (\bibinfo{year}{1930}), ISSN \bibinfo{issn}{0044-3328},
  \urlprefix\url{http://dx.doi.org/10.1007/BF01339933}.

\bibitem[{\citenamefont{Witzel et~al.}(2007)\citenamefont{Witzel, Hu, and
  Das~Sarma}}]{Witzel2007}
\bibinfo{author}{\bibfnamefont{W.~M.} \bibnamefont{Witzel}},
  \bibinfo{author}{\bibfnamefont{X.}~\bibnamefont{Hu}}, \bibnamefont{and}
  \bibinfo{author}{\bibfnamefont{S.}~\bibnamefont{Das~Sarma}},
  \bibinfo{journal}{\prb} \textbf{\bibinfo{volume}{76}},
  \bibinfo{pages}{035212} (\bibinfo{year}{2007}),
  \urlprefix\url{http://link.aps.org/doi/10.1103/PhysRevB.76.035212}.

\bibitem[{\citenamefont{Saikin and Fedichkin}(2003)}]{saikin2003nonideality}
\bibinfo{author}{\bibfnamefont{S.}~\bibnamefont{Saikin}} \bibnamefont{and}
  \bibinfo{author}{\bibfnamefont{L.}~\bibnamefont{Fedichkin}},
  \bibinfo{journal}{Phys. Rev. B} \textbf{\bibinfo{volume}{67}},
  \bibinfo{pages}{161302} (\bibinfo{year}{2003}),
  \urlprefix\url{http://link.aps.org/doi/10.1103/PhysRevB.67.161302}.

\bibitem[{\citenamefont{Burum}(1981)}]{burum1981magnus}
\bibinfo{author}{\bibfnamefont{D.~P.} \bibnamefont{Burum}},
  \bibinfo{journal}{\prb} \textbf{\bibinfo{volume}{24}}, \bibinfo{pages}{3684}
  (\bibinfo{year}{1981}).

\bibitem[{\citenamefont{Maricq}(1982)}]{PhysRevB.25.6622}
\bibinfo{author}{\bibfnamefont{M.~M.} \bibnamefont{Maricq}},
  \bibinfo{journal}{Phys. Rev. B} \textbf{\bibinfo{volume}{25}},
  \bibinfo{pages}{6622} (\bibinfo{year}{1982}),
  \urlprefix\url{http://link.aps.org/doi/10.1103/PhysRevB.25.6622}.

\bibitem[{\citenamefont{{Hung} et~al.}(2013)\citenamefont{{Hung},
  {Cywi{\'n}ski}, {Hu}, and {Das Sarma}}}]{hung2013hyperfine}
\bibinfo{author}{\bibfnamefont{J.-T.} \bibnamefont{{Hung}}},
  \bibinfo{author}{\bibfnamefont{{\L}.}~\bibnamefont{{Cywi{\'n}ski}}},
  \bibinfo{author}{\bibfnamefont{X.}~\bibnamefont{{Hu}}}, \bibnamefont{and}
  \bibinfo{author}{\bibfnamefont{S.}~\bibnamefont{{Das Sarma}}},
  \bibinfo{journal}{ArXiv e-prints}  (\bibinfo{year}{2013}),
  \eprint{1304.6711},
  \urlprefix\url{http://adsabs.harvard.edu/abs/2013arXiv1304.6711H}.

\bibitem[{\citenamefont{Brunner et~al.}(2011)\citenamefont{Brunner, Shin,
  Obata, Pioro-Ladri\`ere, Kubo, Yoshida, Taniyama, Tokura, and
  Tarucha}}]{PhysRevLett.107.146801}
\bibinfo{author}{\bibfnamefont{R.}~\bibnamefont{Brunner}},
  \bibinfo{author}{\bibfnamefont{Y.-S.} \bibnamefont{Shin}},
  \bibinfo{author}{\bibfnamefont{T.}~\bibnamefont{Obata}},
  \bibinfo{author}{\bibfnamefont{M.}~\bibnamefont{Pioro-Ladri\`ere}},
  \bibinfo{author}{\bibfnamefont{T.}~\bibnamefont{Kubo}},
  \bibinfo{author}{\bibfnamefont{K.}~\bibnamefont{Yoshida}},
  \bibinfo{author}{\bibfnamefont{T.}~\bibnamefont{Taniyama}},
  \bibinfo{author}{\bibfnamefont{Y.}~\bibnamefont{Tokura}}, \bibnamefont{and}
  \bibinfo{author}{\bibfnamefont{S.}~\bibnamefont{Tarucha}},
  \bibinfo{journal}{Phys. Rev. Lett.} \textbf{\bibinfo{volume}{107}},
  \bibinfo{pages}{146801} (\bibinfo{year}{2011}),
  \urlprefix\url{http://link.aps.org/doi/10.1103/PhysRevLett.107.146801}.

\bibitem[{\citenamefont{Coish and Baugh}(2009)}]{PSSB:PSSB200945229}
\bibinfo{author}{\bibfnamefont{W.~A.} \bibnamefont{Coish}} \bibnamefont{and}
  \bibinfo{author}{\bibfnamefont{J.}~\bibnamefont{Baugh}},
  \bibinfo{journal}{{P}hysica {S}tatus {S}olidi (b)}
  \textbf{\bibinfo{volume}{246}}, \bibinfo{pages}{2203} (\bibinfo{year}{2009}),
  ISSN \bibinfo{issn}{1521-3951},
  \urlprefix\url{http://dx.doi.org/10.1002/pssb.200945229}.

\bibitem[{\citenamefont{Assali et~al.}(2011)\citenamefont{Assali, Petrilli,
  Capaz, Koiller, Hu, and Das~Sarma}}]{PhysRevB.83.165301}
\bibinfo{author}{\bibfnamefont{L.~V.~C.} \bibnamefont{Assali}},
  \bibinfo{author}{\bibfnamefont{H.~M.} \bibnamefont{Petrilli}},
  \bibinfo{author}{\bibfnamefont{R.~B.} \bibnamefont{Capaz}},
  \bibinfo{author}{\bibfnamefont{B.}~\bibnamefont{Koiller}},
  \bibinfo{author}{\bibfnamefont{X.}~\bibnamefont{Hu}}, \bibnamefont{and}
  \bibinfo{author}{\bibfnamefont{S.}~\bibnamefont{Das~Sarma}},
  \bibinfo{journal}{Phys. Rev. B} \textbf{\bibinfo{volume}{83}},
  \bibinfo{pages}{165301} (\bibinfo{year}{2011}),
  \urlprefix\url{http://link.aps.org/doi/10.1103/PhysRevB.83.165301}.

\bibitem[{\citenamefont{Petta et~al.}(2005)\citenamefont{Petta, Johnson,
  Taylor, Laird, Yacoby, Lukin, Marcus, Hanson, and
  Gossard}}]{petta2005coherent}
\bibinfo{author}{\bibfnamefont{J.}~\bibnamefont{Petta}},
  \bibinfo{author}{\bibfnamefont{A.}~\bibnamefont{Johnson}},
  \bibinfo{author}{\bibfnamefont{J.}~\bibnamefont{Taylor}},
  \bibinfo{author}{\bibfnamefont{E.}~\bibnamefont{Laird}},
  \bibinfo{author}{\bibfnamefont{A.}~\bibnamefont{Yacoby}},
  \bibinfo{author}{\bibfnamefont{M.}~\bibnamefont{Lukin}},
  \bibinfo{author}{\bibfnamefont{C.}~\bibnamefont{Marcus}},
  \bibinfo{author}{\bibfnamefont{M.}~\bibnamefont{Hanson}}, \bibnamefont{and}
  \bibinfo{author}{\bibfnamefont{A.}~\bibnamefont{Gossard}},
  \bibinfo{journal}{Science} \textbf{\bibinfo{volume}{309}},
  \bibinfo{pages}{2180} (\bibinfo{year}{2005}).

\bibitem[{\citenamefont{Schliemann et~al.}(2003)\citenamefont{Schliemann,
  Khaetskii, and Loss}}]{schliemann2003electron}
\bibinfo{author}{\bibfnamefont{J.}~\bibnamefont{Schliemann}},
  \bibinfo{author}{\bibfnamefont{A.}~\bibnamefont{Khaetskii}},
  \bibnamefont{and} \bibinfo{author}{\bibfnamefont{D.}~\bibnamefont{Loss}},
  \bibinfo{journal}{Journal of Physics: Condensed Matter}
  \textbf{\bibinfo{volume}{15}}, \bibinfo{pages}{R1809} (\bibinfo{year}{2003}).

\bibitem[{\citenamefont{Witzel et~al.}(2012)\citenamefont{Witzel, Carroll,
  Cywi\ifmmode~\acute{n}\else \'{n}\fi{}ski, and
  Das~Sarma}}]{PhysRevB.86.035452}
\bibinfo{author}{\bibfnamefont{W.~M.} \bibnamefont{Witzel}},
  \bibinfo{author}{\bibfnamefont{M.~S.} \bibnamefont{Carroll}},
  \bibinfo{author}{\bibfnamefont{L.}~\bibnamefont{Cywi\ifmmode~\acute{n}\else
  \'{n}\fi{}ski}}, \bibnamefont{and}
  \bibinfo{author}{\bibfnamefont{S.}~\bibnamefont{Das~Sarma}},
  \bibinfo{journal}{Phys. Rev. B} \textbf{\bibinfo{volume}{86}},
  \bibinfo{pages}{035452} (\bibinfo{year}{2012}),
  \urlprefix\url{http://link.aps.org/doi/10.1103/PhysRevB.86.035452}.

\bibitem[{\citenamefont{Tyryshkin et~al.}(2003)\citenamefont{Tyryshkin, Lyon,
  Astashkin, and Raitsimring}}]{PhysRevB.68.193207}
\bibinfo{author}{\bibfnamefont{A.~M.} \bibnamefont{Tyryshkin}},
  \bibinfo{author}{\bibfnamefont{S.~A.} \bibnamefont{Lyon}},
  \bibinfo{author}{\bibfnamefont{A.~V.} \bibnamefont{Astashkin}},
  \bibnamefont{and} \bibinfo{author}{\bibfnamefont{A.~M.}
  \bibnamefont{Raitsimring}}, \bibinfo{journal}{Phys. Rev. B}
  \textbf{\bibinfo{volume}{68}}, \bibinfo{pages}{193207}
  (\bibinfo{year}{2003}),
  \urlprefix\url{http://link.aps.org/doi/10.1103/PhysRevB.68.193207}.

\bibitem[{\citenamefont{Abe et~al.}(2010)\citenamefont{Abe, Tyryshkin, Tojo,
  Morton, Witzel, Fujimoto, Ager, Haller, Isoya, Lyon
  et~al.}}]{PhysRevB.82.121201}
\bibinfo{author}{\bibfnamefont{E.}~\bibnamefont{Abe}},
  \bibinfo{author}{\bibfnamefont{A.~M.} \bibnamefont{Tyryshkin}},
  \bibinfo{author}{\bibfnamefont{S.}~\bibnamefont{Tojo}},
  \bibinfo{author}{\bibfnamefont{J.~J.~L.} \bibnamefont{Morton}},
  \bibinfo{author}{\bibfnamefont{W.~M.} \bibnamefont{Witzel}},
  \bibinfo{author}{\bibfnamefont{A.}~\bibnamefont{Fujimoto}},
  \bibinfo{author}{\bibfnamefont{J.~W.} \bibnamefont{Ager}},
  \bibinfo{author}{\bibfnamefont{E.~E.} \bibnamefont{Haller}},
  \bibinfo{author}{\bibfnamefont{J.}~\bibnamefont{Isoya}},
  \bibinfo{author}{\bibfnamefont{S.~A.} \bibnamefont{Lyon}},
  \bibnamefont{et~al.}, \bibinfo{journal}{Phys. Rev. B}
  \textbf{\bibinfo{volume}{82}}, \bibinfo{pages}{121201}
  (\bibinfo{year}{2010}),
  \urlprefix\url{http://link.aps.org/doi/10.1103/PhysRevB.82.121201}.

\bibitem[{\citenamefont{Kane}(1998)}]{kane1998silicon}
\bibinfo{author}{\bibfnamefont{B.}~\bibnamefont{Kane}},
  \bibinfo{journal}{Nature} \textbf{\bibinfo{volume}{393}},
  \bibinfo{pages}{133} (\bibinfo{year}{1998}).

\bibitem[{\citenamefont{Coish and Loss}(2005)}]{PhysRevB.72.125337}
\bibinfo{author}{\bibfnamefont{W.~A.} \bibnamefont{Coish}} \bibnamefont{and}
  \bibinfo{author}{\bibfnamefont{D.}~\bibnamefont{Loss}},
  \bibinfo{journal}{Phys. Rev. B} \textbf{\bibinfo{volume}{72}},
  \bibinfo{pages}{125337} (\bibinfo{year}{2005}),
  \urlprefix\url{http://link.aps.org/doi/10.1103/PhysRevB.72.125337}.

\bibitem[{\citenamefont{Paget et~al.}(1977)\citenamefont{Paget, Lampel,
  Sapoval, and Safarov}}]{Paget1977low}
\bibinfo{author}{\bibfnamefont{D.}~\bibnamefont{Paget}},
  \bibinfo{author}{\bibfnamefont{G.}~\bibnamefont{Lampel}},
  \bibinfo{author}{\bibfnamefont{B.}~\bibnamefont{Sapoval}}, \bibnamefont{and}
  \bibinfo{author}{\bibfnamefont{V.}~\bibnamefont{Safarov}},
  \bibinfo{journal}{\prb} \textbf{\bibinfo{volume}{15}}, \bibinfo{pages}{5780}
  (\bibinfo{year}{1977}).

\end{thebibliography}

\end{document}